\documentclass{aa}  

\usepackage{graphicx}
\usepackage{txfonts}

\usepackage{hyperref}

\makeatletter
\renewcommand*\aa@pageof{, page \thepage{} of \pageref*{LastPage}}
\makeatother

\def\ltsima{$\; \buildrel < \over \sim \;$}
\def\simlt{\lower.5ex\hbox{\ltsima}}
\def\gtsima{$\; \buildrel > \over \sim \;$}
\def\simgt{\lower.5ex\hbox{\gtsima}}
\def\gsimeq
{\hbox{\raise0.5ex\hbox{$>\lower1.06ex\hbox{$\kern-1.07em{\sim}$}$}}}
\def\lsimeq
{\hbox{\raise0.5ex\hbox{$<\lower1.06ex\hbox{$\kern-1.07em{\sim}$}$}}}

\def\rosat{{\it ROSAT}}

\def\erosita{{\it eROSITA}}

\def\aap{A\&A}

\def\apjl{ApJ}
\def\apjs{ApJS}

\def\erosita{{\it eROSITA}}

\begin{document}

   \title{Low-mass stars dominate the hot (0.7~keV) Galactic X-ray emission}

   \author{G. Ponti\inst{1,2,3}
          \and
          M. C. H. Yeung\inst{2}
          \and
          G. Stel\inst{1,3}
          \and
          N. Locatelli\inst{1}
          \and
            X. Zheng\inst{2}
          \and
            B. Stelzer\inst{4}
          \and
            A. Merloni\inst{2}
          \and
            M. Caramazza\inst{4}
          \and
            E. Magaudda\inst{4}
          \and
            M. Sasaki\inst{5}
          \and
            K. Dennerl\inst{2}
          \and
            T. H. Reiprich\inst{6}
          \and
            A. Schwope\inst{7}
          \and
            W. Becker\inst{2}
          \and
            M. Freyberg\inst{2}          
          }

   \institute{Osservatorio Astronomico di Brera, (INAF), Via E. Bianchi 46, Merate, 23807, Italy
              \email{gabriele.ponti@inaf.it}
          \and
Max-Planck-Institut f\"ur extraterrestrische Physik, Giessenbachstrasse, Garching, 85748, Germany
          \and
Como Lake Center for Astrophysics (CLAP), DiSAT, Università degli Studi dell’Insubria, via Valleggio 11, 22100 Como, Italy
          \and
Institut f\"ur Astronomie \& Astrophysik, Eberhard Karls Universit\"at T\"ubingen, Sand 1, T\"ubingen, 72076, Germany
          \and
Dr. Karl Remeis Observatory, Erlangen Centre for Astroparticle Physics, Friedrich-Alexander-Universität Erlangen-Nürnberg, Sternwartstr. 7, Bamberg, 96049, Germany
          \and
Argelander Institute for Astronomy, University of Bonn, Auf dem H\"ugel 71, Bonn, 53121, Germany
          \and
Leibniz-Institut f\"ur Astrophysik Potsdam (AIP), An der Sternwarte 16, Potsdam, 14482, Germany
             }

   \date{Received 26 August 2025; accepted XXX}

  \abstract
   {The circumgalactic medium (CGM) of the Milky Way is composed of a tenuous atmosphere filled with multi-phase plasma, including a warm-hot virialised component. Recent studies suggest a much hotter ($\sim$0.7\,keV) super-virial component detected in both absorption and emission.}
   {We want to shed light on the nature of this putative super-virial component.}
   {We analysed the X-ray background as observed by {\it SRG}/\erosita\ over the entire western Galactic hemisphere.}
   {We show that low-mass stars provide a large fraction of the 0.7\,keV emission. 
Indeed, a tight correlation is found between the surface brightness of the 0.7~keV emission and the mass distribution of the Milky Way across a large portion of the western Galactic hemisphere.
The correlation coefficient implies an X-ray luminosity per unit of stellar mass comparable to that of the average low-mass stars within 10~pc of the Sun, suggesting that unresolved M dwarfs and F, G, and K type stars dominate the 0.7~keV emission. 
This emission is asymmetric with respect to the Galactic plane, influenced by the asymmetric distribution of nearby star-forming regions, and broadly consistent with the known offset of the Sun above the Galactic midplane.
The remaining signal might be produced by the cumulative emission of stars of different types or ages, in addition to other sources (e.g. hot interstellar medium, Galactic corona, etc.). Assuming that the putative residual hot super-virial atmosphere is homogeneous and has a spherical beta profile with slope $\beta=0.4$, we constrain its density at 10 kpc to be $n_e<4\times10^{-4}$~cm$^{-3}$. Our findings may help refine models of the circumgalactic medium around external galaxies, advancing our understanding of hot baryon flows and galaxy evolution.}
   {}


   \maketitle
%

\section{Introduction}

Standard galaxy formation theory predicts a tenuous atmosphere around Milky Way-like galaxies, which are bound by their dark-matter halos \citep{White91}. 
This plasma is expected to be virialised, have temperatures of a few million degrees ($kT\sim0.1-0.2$ keV), and be emitting X-rays \citep{Tumlinson17}. 

Our Milky Way exhibits a warm-hot virialised atmosphere, but some studies also suggested that the circumgalactic medium (CGM) might exhibit multiple temperatures \citep{Yao07,Yao09} and include a much hotter ($\sim$0.7\,keV) super-virial component recently detected in both absorption and emission \citep{Yoshino09,Miller15,Nicastro18,Kaaret19,Das19a,Gupta21,Ponti23}. 
In fact, cosmological simulations indicate that a fraction of the CGM plasma might be heated to super-virial temperatures \citep{Ramesh23}, suggesting that a super-virial CGM component might be present around the Milky Way. 
However, the 0.7~keV gas seen in emission cannot explain the absorption observations \citep{Bisht24ApJ}, possibly indicating different origins. 

Low-mass stars of spectral types F, G, K, and M, which contribute substantially to the stellar mass of the Milky Way, are likely responsible for part of the 0.7\,keV emission due to their X-ray-bright coronae  \citep{Guedel04,Magaudda22,Caramazza23}. 
Indeed, studies of the soft X-ray background at low Galactic latitudes have suggested that active stars could contribute close to the Galactic plane \citep{Rosner81,Masui09,Wulf19,Ampuku24}, while their contribution at mid and high Galactic latitudes was unclear. 
In particular, \citet{Wulf19} analysed four calorimetric observations from sounding rocket experiments and suggested that stars, in particular young M dwarfs, contribute significantly to the 0.7\,keV emission at low Galactic latitudes. In this work, we studied the soft X-ray background observed by SRG/\erosita\ \citep{Sunyaev21,Predehl21,Merloni24},
to shed light on the nature of the 0.7~keV emission.

\section{Spectral analysis of the soft X-ray background}
\label{sec:analysis}

\begin{figure*}[t]
\centering
\includegraphics[width=0.47\textwidth]{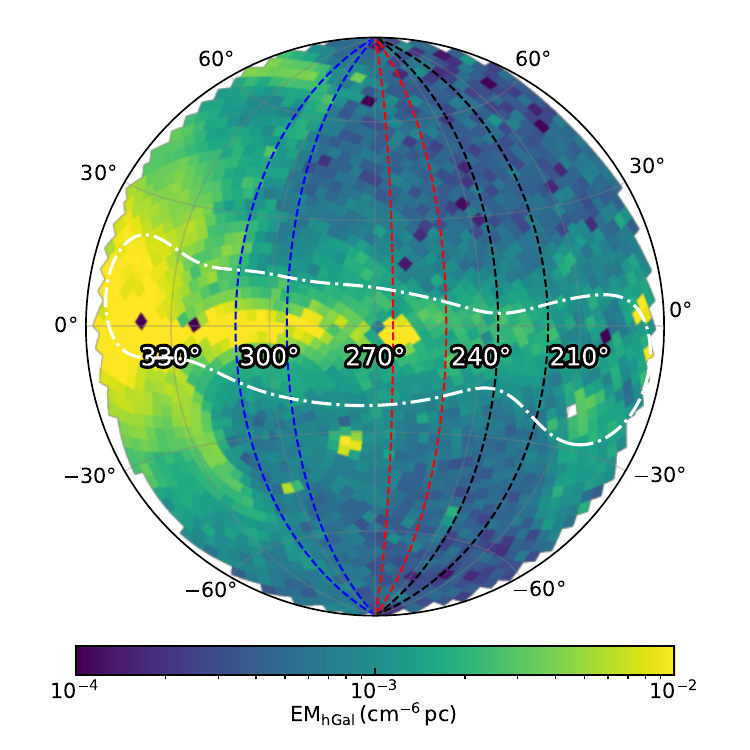}
\includegraphics[width=0.47\textwidth]{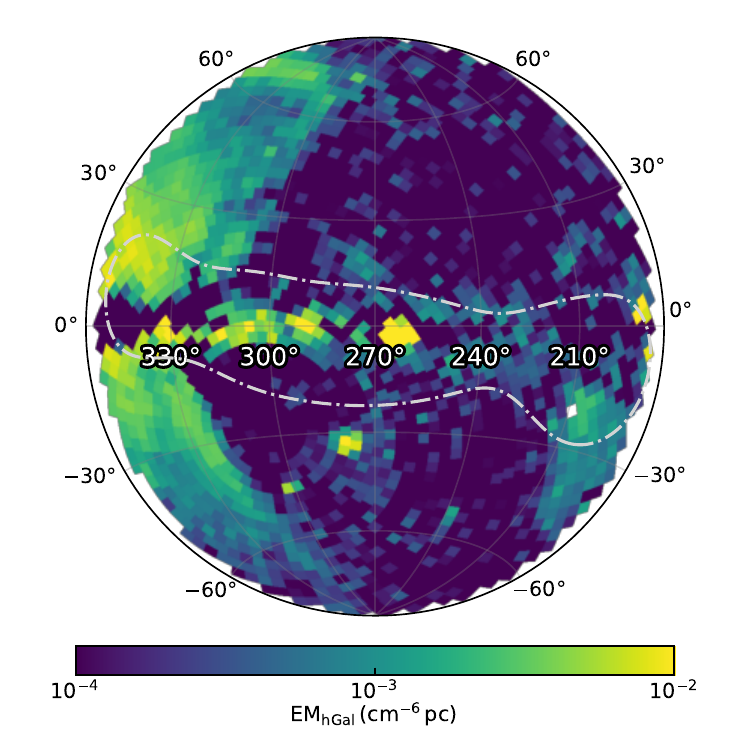}
\caption{{\it Left panel:} Logarithm of emission measure of the 0.7\,keV Galactic emission as observed by \erosita\ in the western Galactic hemisphere when fitting the total X-ray emission (diffuse and point sources); it is given in Galactic coordinates and in zenithal equal-area projection centred on ($l=270^\circ$,~$b=0^\circ$). The dashed black, red, and blue lines show the longitudinal stripes in Fig. \ref{fig:Multistripe}. The dot-dashed white contours indicate the region with the highest concentration of stars within 500~pc of the Sun, as shown in Fig. \ref{fig:StarFrac}. 
{\it Right panel:} Same as left panel, but once the stellar contribution, as described by \cite{Hunter24}, had been removed.} 
\label{fig:MapCoro}
\end{figure*}

\begin{figure*}[t]
\centering
\vspace*{-0.8cm}
\includegraphics[width=0.45\textwidth]{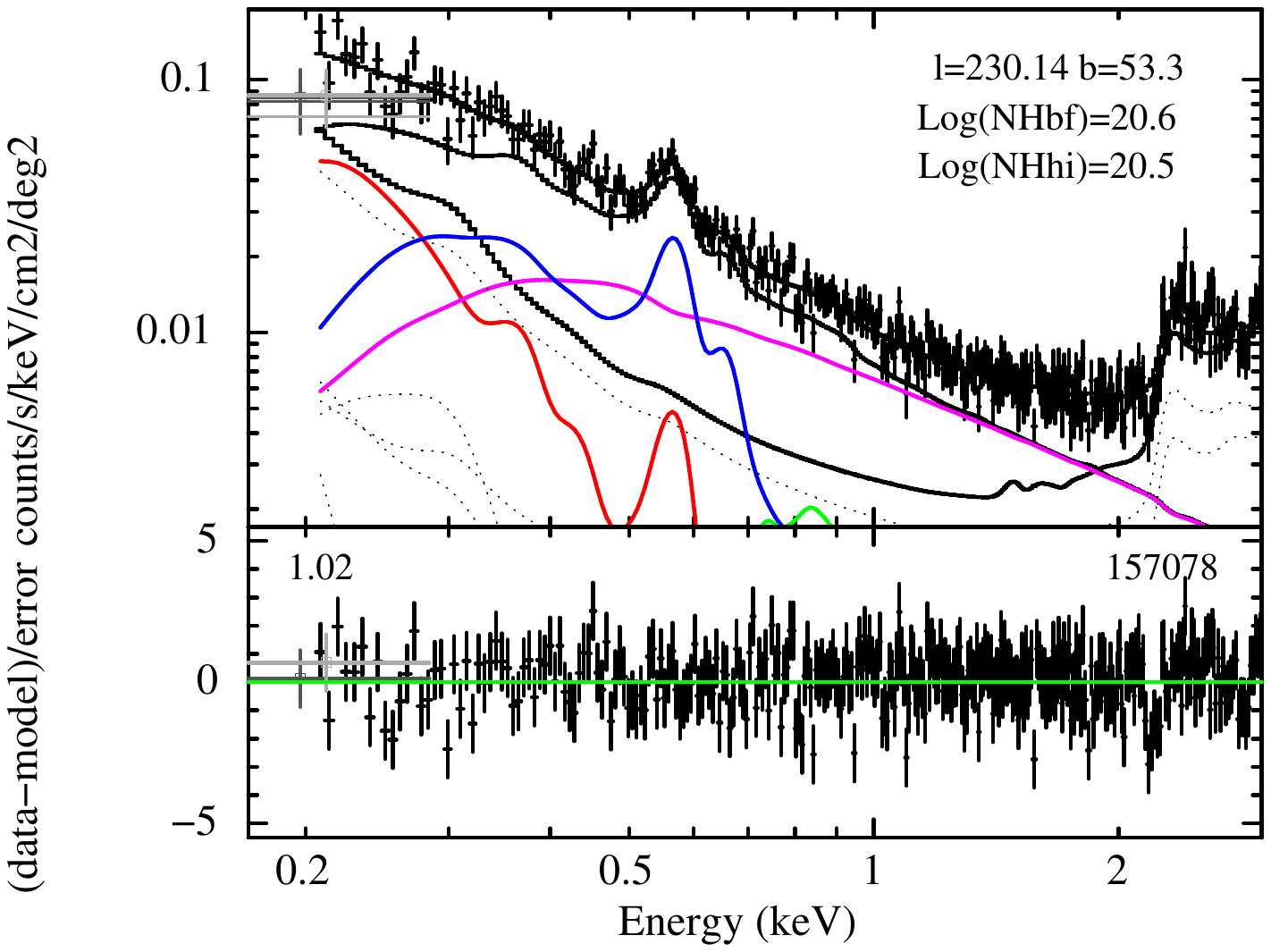}
\hspace*{-0.5cm}
\includegraphics[width=0.45\textwidth]{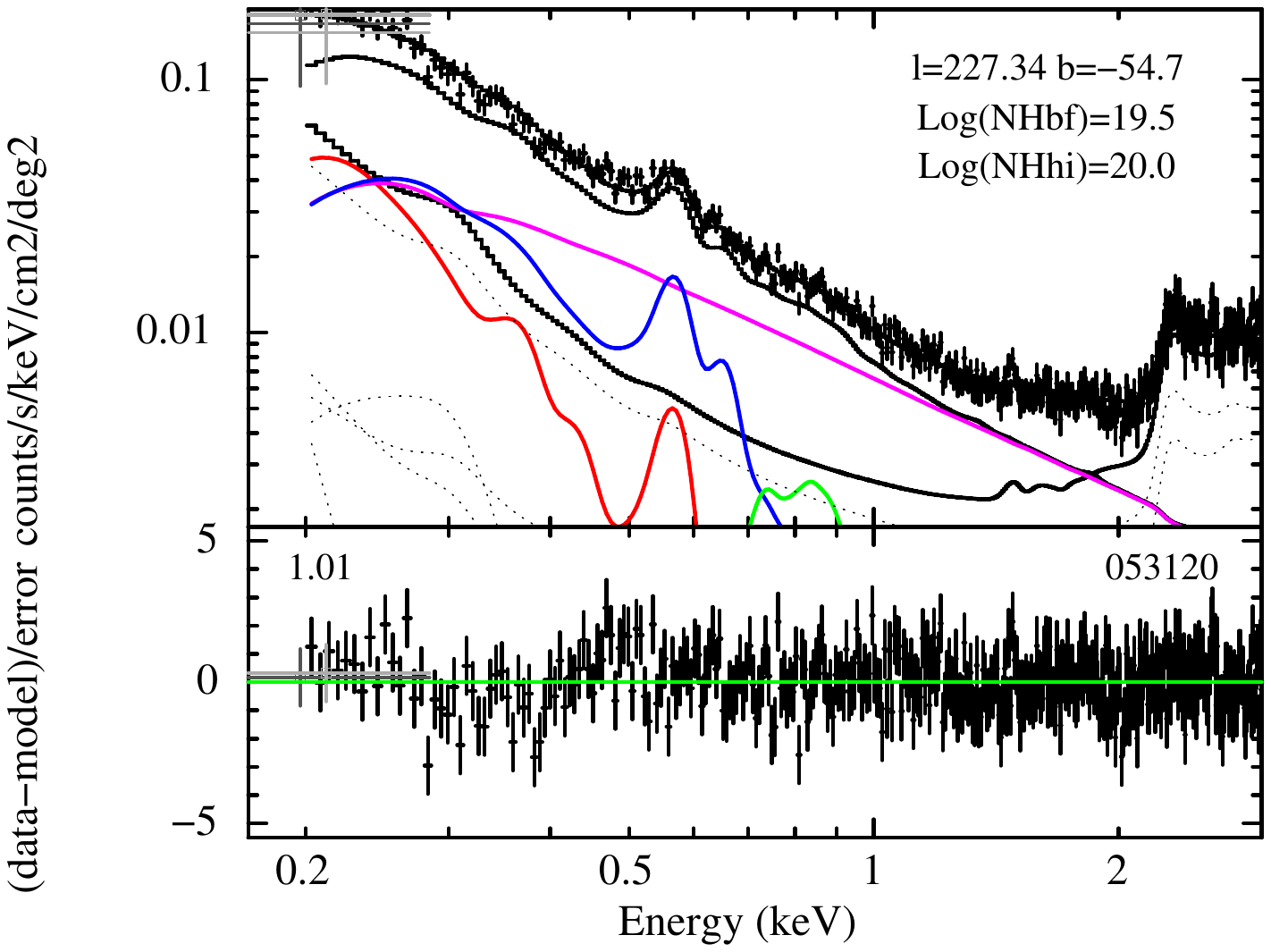}
\vspace*{-1.1cm}

\includegraphics[width=0.45\textwidth]{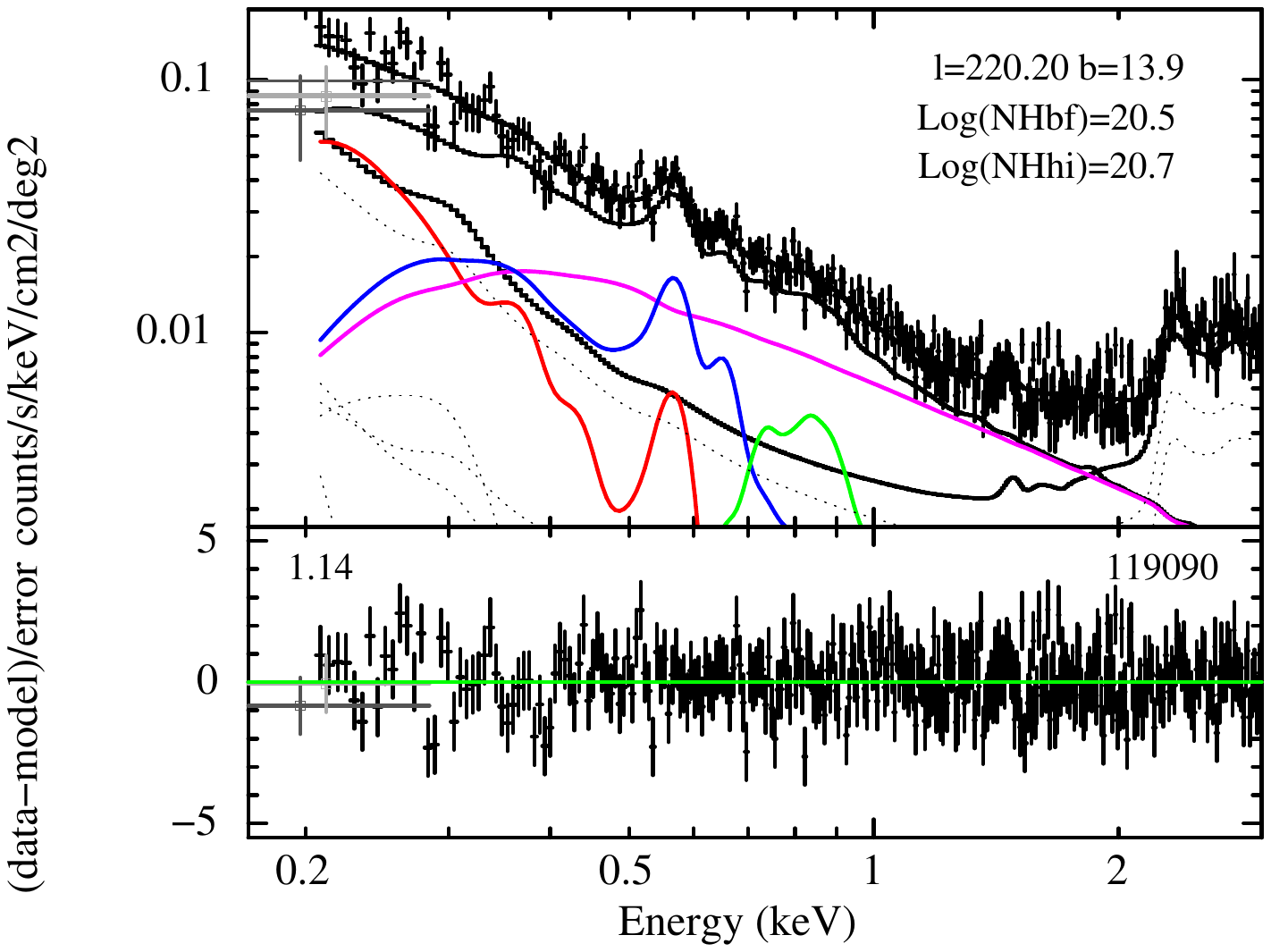}
\hspace*{-0.5cm}
\includegraphics[width=0.45\textwidth]{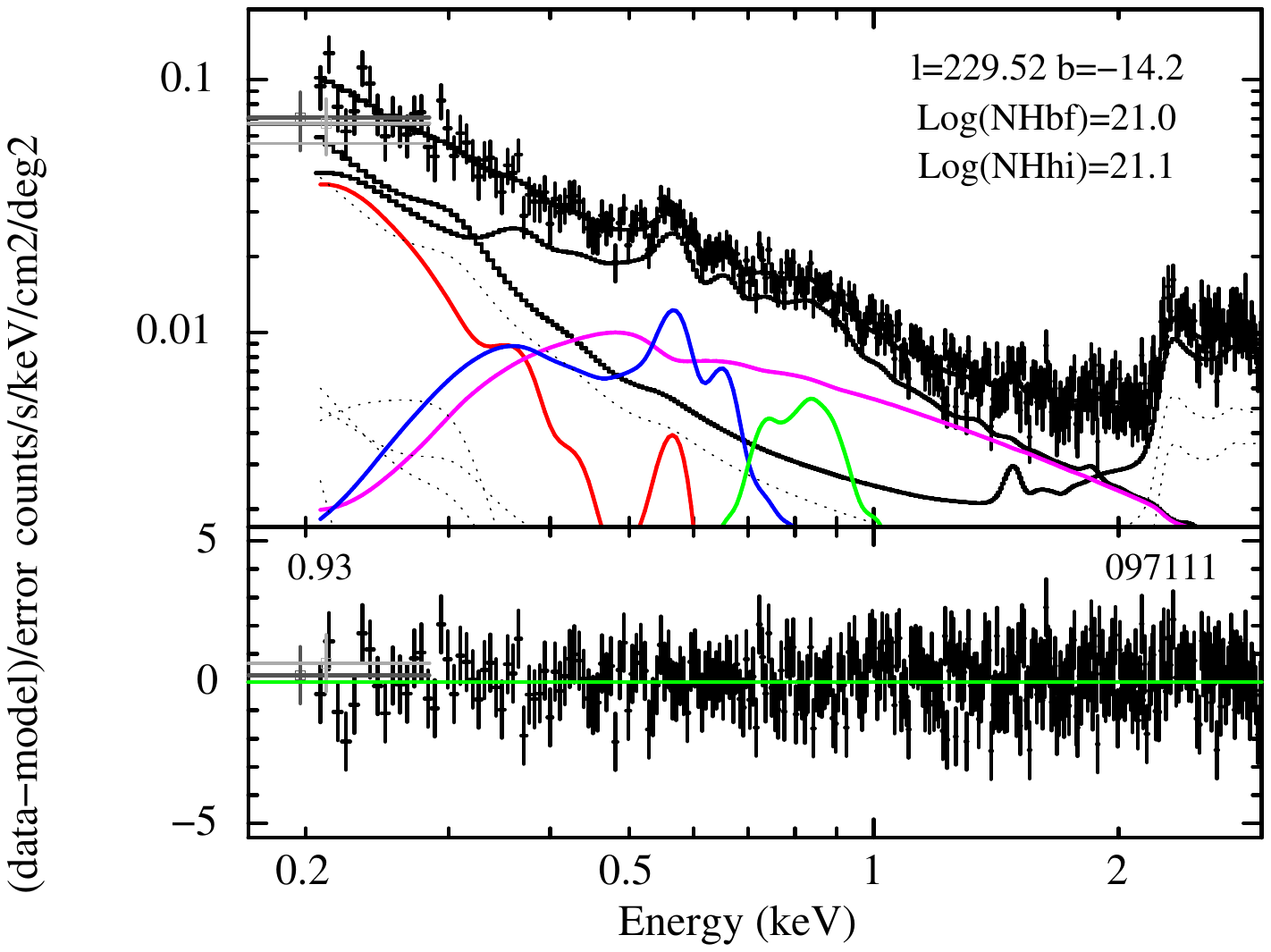}
\vspace*{-1.1cm}

\includegraphics[width=0.45\textwidth]{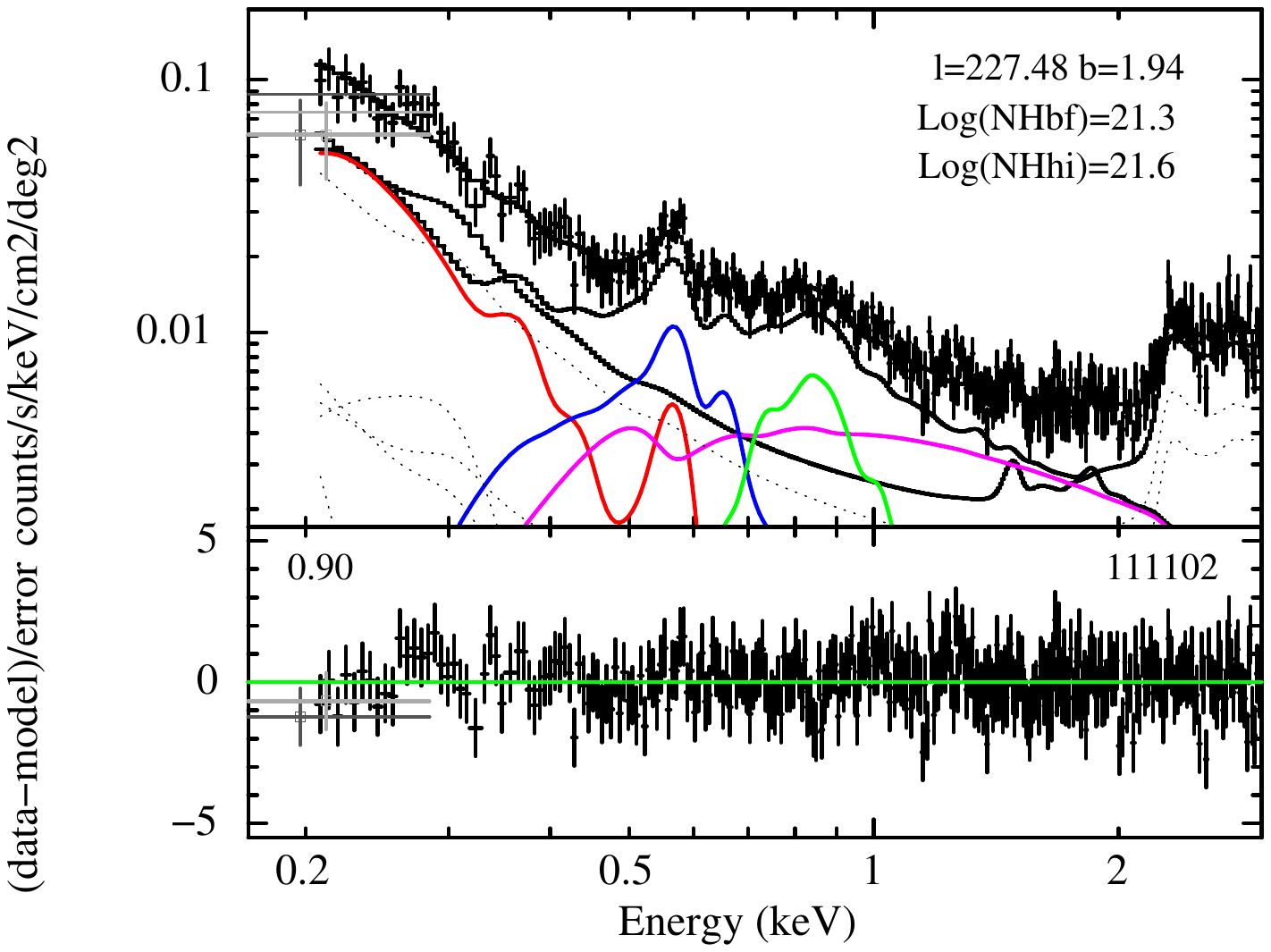}
\hspace*{-0.5cm}
\includegraphics[width=0.45\textwidth]{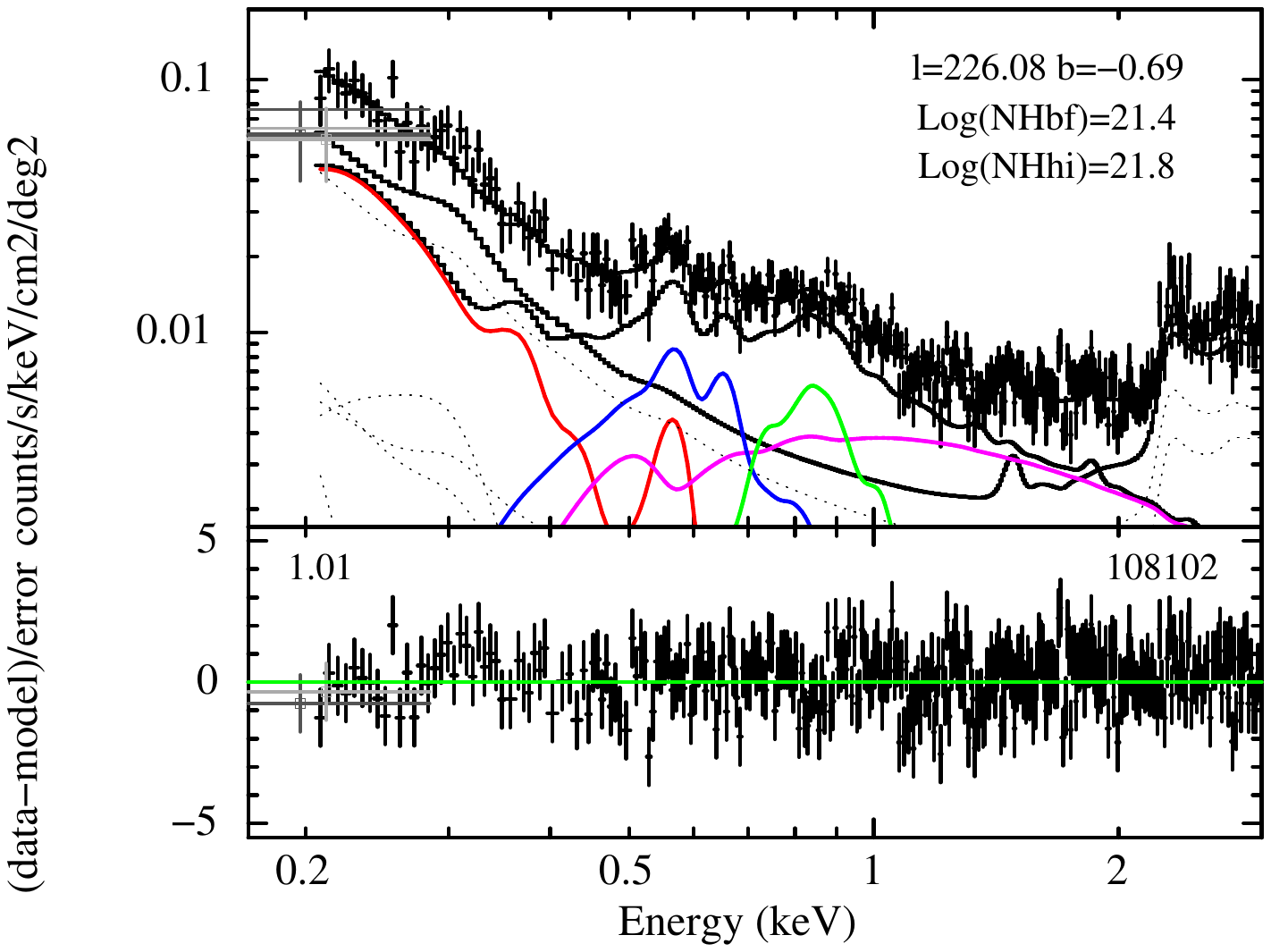}
\vspace*{-0.4cm}
\caption{Best-fit spectra for six selected sky tiles within $220^\circ<l<235^\circ$, within the dashed black lines shown in Fig. \ref{fig:MapCoro}. 
The left column shows spectra within the northern hemisphere at decreasing latitudes. 
The right column shows spectra at similar latitudes in the southern hemisphere. 
The black data show the \erosita\ spectrum, while the grey points show the flux measured by \rosat\ in the softest bands. 
The solid lines show the contribution from the local hot bubble (red), the warm-hot bubble (blue), the 0.7~keV component (green), the cosmic X-ray background (CXB; magenta), and the instrumental background (black). 
The inset reports the Galactic latitude and longitude of the centre of the sky tile as well as the total hydrogen column density of neutral absorption derived from the HI4PI survey $N_{\rm H,HI4}$ and the best-fit value, $N_{\rm H,bf}$. 
The dotted lines show the contribution of the various components of the instrumental background. 
The lower panels show the fit residuals, with the sky tile number and reduced $\chi^2$.
} 
\label{fig:spec}
\end{figure*}

We analysed the soft X-ray emission over large parts of the sky as observed by \erosita\ (see Appendix \ref{sec:ana} for more details). We divided the western Galactic hemisphere into $\sim3^\circ\times3^\circ$ sky tiles \citep{Merloni24}.
For each tile, we extracted a spectrum of the total X-ray emission (point sources and diffuse) and two data points from the softest bands of the ROSAT all-sky maps. 
We fitted each tile (see Figs. \ref{fig:MapCoro} and \ref{fig:spec}) with the spectral model described by \cite{Ponti23} and \cite{Yeung24} including the cosmic X-ray background (CXB) with a fixed slope and free normalisation (norm$_{\rm CXB}$), two thermal components for the warm-hot (wh) and hot (0.7\,keV) Galactic emission (hGal), and absorption from a single neutral layer fitted with the {\sc disnht} model \citep{Locatelli22} leaving free the column density of neutral absorption ($\log (N_{\rm H})$; see Appendix \ref{sec:absorption}). 
The first thermal component is likely linked to the virialised plasma in the Milky Way halo; therefore, we assumed abundances of $Z=0.1$~Z$_\odot$ \citep{Ponti23} and fitted for the temperature ($kT_{\rm wh}$) and emission measure (${\rm EM_{wh}}$), while the hotter component is the focus of this study. 
The abundance and temperature of the 0.7\,keV component were fixed to solar values and $kT_{hGal}=0.7$\,keV, respectively, and we fitted for the emission measure (${\rm EM_{hGal}}$). 
We also included emission from the local hot bubble (LHB; \cite{Yeung24}) as an unabsorbed component with $kT=0.099$\,keV and solar abundances, and we fitted for the emission measure (${\rm EM_{LHB}}$).
To minimise absorption effects, we only considered sky tiles with total hydrogen column densities of $\log (N_{\rm H}/{\rm cm^{-2}})<21.5$. In fact, the single absorption layer assumed in this spectral model is likely not a good approximation at higher column densities. For example, at high column densities along the Galactic disc, the cold and hot phases are expected to be mixed, generating a more complex absorption pattern than the single-layer scenario assumed here (see Appendix \ref{sec:absorption}). 
The model includes six free parameters: ${\rm EM_{LHB}}$, $\log{N_{\rm H,bf}}$, $kT_{\rm wh}$, ${\rm EM_{wh}}$, ${\rm EM_{hGal}}$, and ${\rm norm_{CXB}}$; and in {\sc Xspec} terminology is written as {\sc apec$_{\rm LHB}$ + disnht (bkn2pow + apec$_{\rm WH}$ + apec$_{\rm hGal}$)}, where {\sc apec}$_{\rm LHB}$, {\sc apec}$_{\rm WH}$, and {\sc apec}$_{\rm hGal}$ represent the thermal components for the local hot bubble, the warm-hot (virial) CGM, and the 0.7~keV emission, respectively, while the {\sc disnht} and the {\sc bkn2pow} represent the neutral absorption from the interstellar medium and the emission from the cosmic X-ray background, respectively.

Figure \ref{fig:spec} shows six examples of spectral fitting in six different sky tiles, all aligned within the longitudinal stripe $220\degr<l<235\degr$, indicating an increase of the 0.7 keV emission component towards low Galactic latitudes.
The left panel of Fig. \ref{fig:MapCoro} shows the logarithmic emission measure (see Appendix \ref{sec:EM} for the definition) of the 0.7\,keV component for each sky tile. 
A clear increase in emission towards the Galactic disc and Galactic centre is observed, with emission enhancements near known bright sources not accounted for in the model. 
To mitigate this, we created a mask to exclude these affected tiles (Appendix \ref{sec:mask} and Fig. \ref{fig:MapChi2}).

\begin{figure}[th]
\centering
\includegraphics[width=0.49\textwidth]{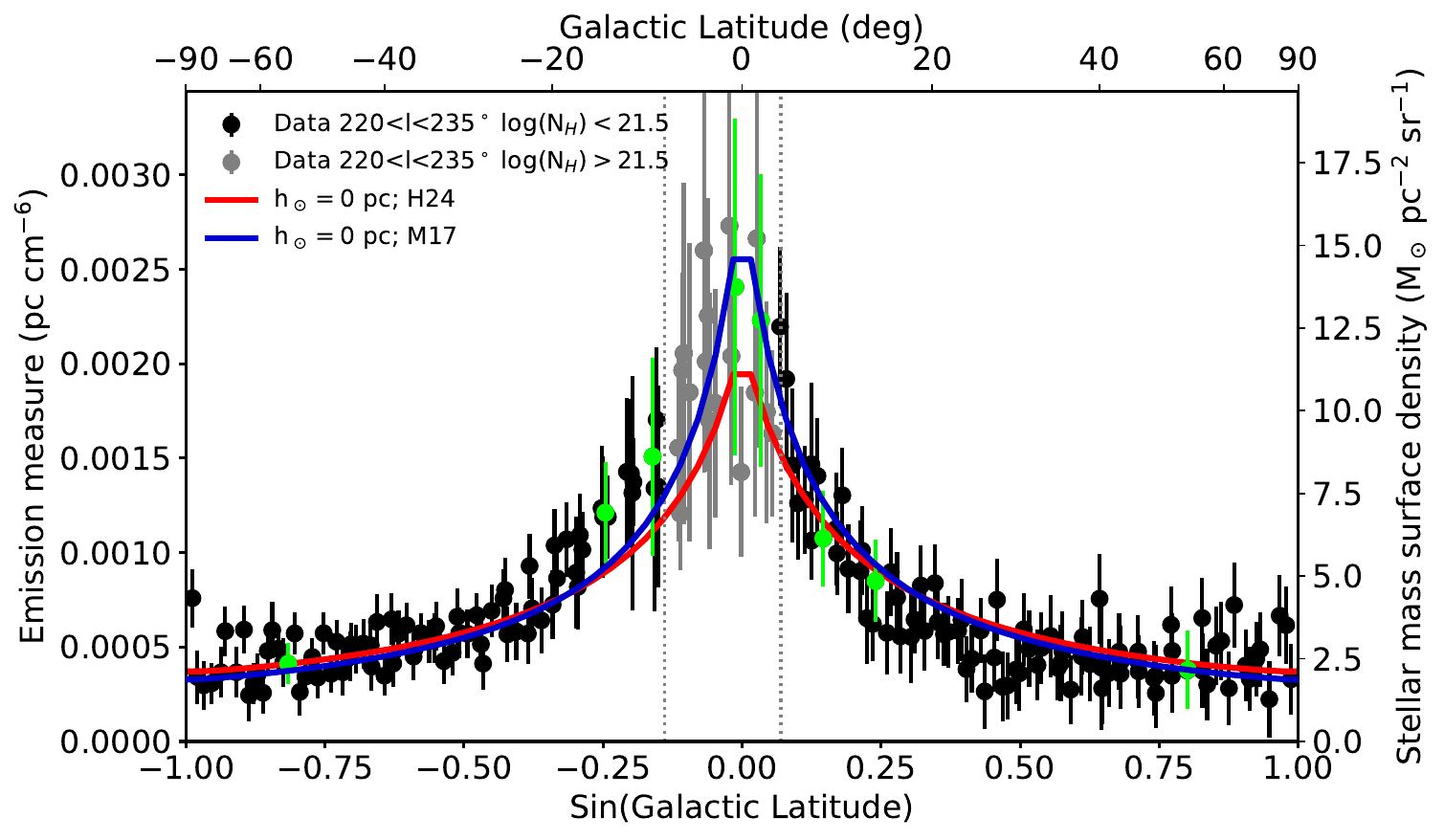}
\includegraphics[width=0.49\textwidth]{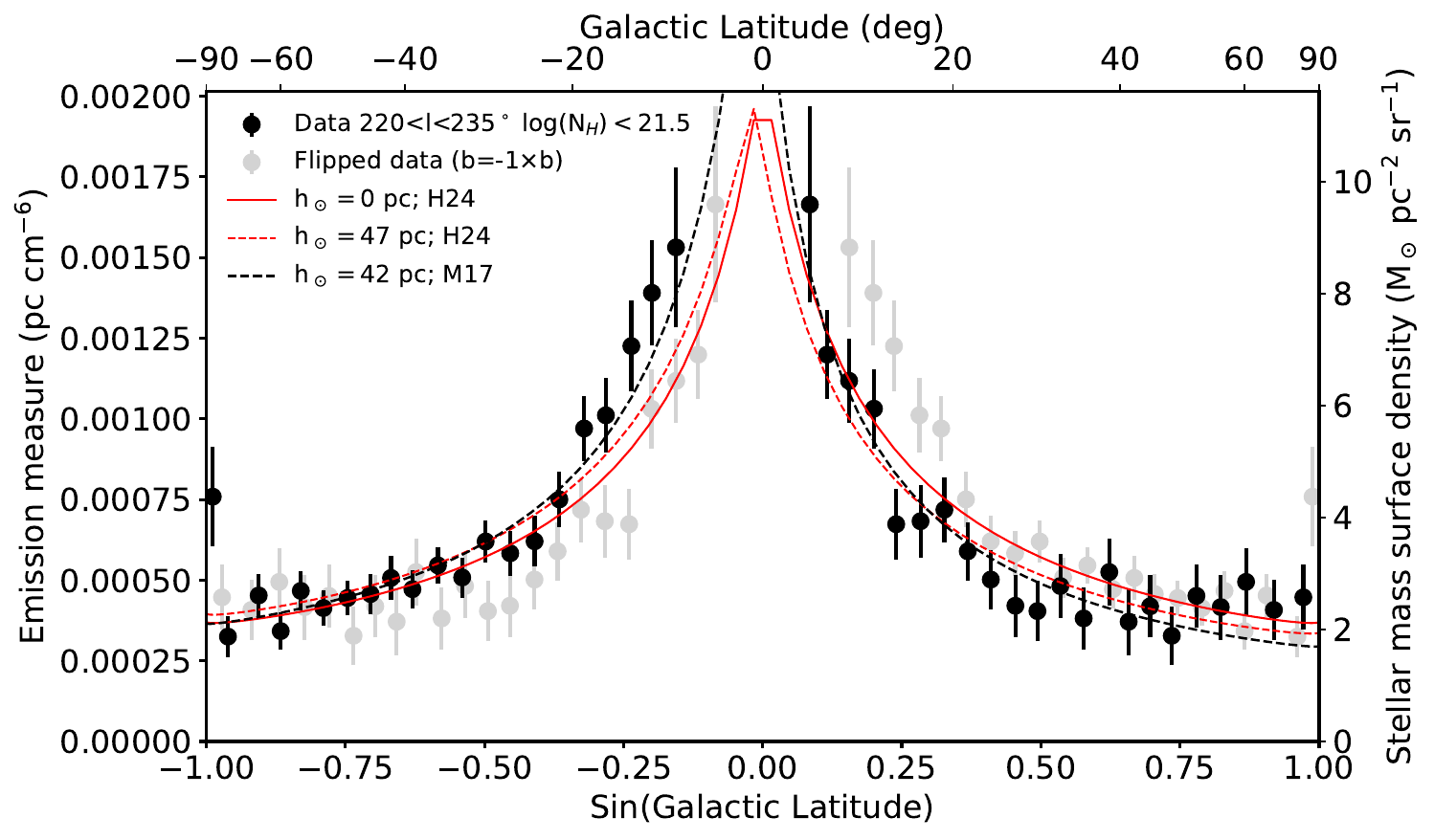}
\caption{Emission measure of 0.7~keV emission for sky tiles within $220^\circ<l<235^\circ$. 
{\it Top panel:} Black and grey data points show the emission measure for the sky tiles with total hydrogen column density of absorption lower and higher than $\log(N_{\rm H,HI4}/{\rm cm^{-2}})>21.5$, respectively. 
The red and blue curves show the best fit with the mass distribution of the Milky Way as described by \cite{Hunter24} and \cite{McMillan17}, respectively, assuming the Sun to lie on the Galactic plane. 
The right y-axis in mass surface density is valid for the \cite{Hunter24} best-fit normalisation. 
The vertical dotted grey lines indicate the regions close to the Galactic disc, where the values from the HI4PI survey are in excess of $\log(N_{\rm H,HI4}/{\rm cm^{-2}})>21.5$. 
The green data show the emission measure of the spectra displayed in Fig. \ref{fig:spec}.
{\it Bottom panel:} Black data show the same emission measures as in the top panel, re-binned over four consecutive (in Galactic latitude) sky tiles. 
The grey points show the same data, once the Galactic latitudes were flipped in terms of the sign (e.g. inverted northern and southern hemispheres). 
The solid and dashed red lines show the best-fit relation when the mass distribution is described by \cite{Hunter24}, with the Sun located on the Galactic plane and 47~pc above it, respectively. 
The dashed black line shows the best-fit mass distribution as described by \cite{McMillan17} for a height of the Sun of 42~pc above the Galactic plane. 
The right y-axis in mass surface density is valid for the \cite{Hunter24} model with $h_\odot=0$ best-fit normalisation. }
\label{fig:Coro}
\end{figure}

The top panel of Fig. \ref{fig:Coro} displays the emission measure of the 0.7 keV component for sky tiles within longitudes $220^\circ<l<235^\circ$ (dashed black lines in Fig. \ref{fig:MapCoro}). 
The black and grey points show the emission measure for sky tiles with total hydrogen column density estimated through the HI4PI survey \citep[$N_{\rm H, HI4}$;][]{HI4PIColl16} smaller and larger than $\log(N_{\rm H, HI4})=21.5$, respectively. 
The 0.7\,keV component reveals variations in the emission measure, which increases closer to the Galactic plane in both hemispheres. 
The red and blue lines in Fig. \ref{fig:Coro} show the stellar mass surface density of the Milky Way, based on models by \cite{Hunter24} and \cite{McMillan17}, respectively, in units of M$_\odot$ pc$^{-2}$ sr$^{-1}$, where M$_\odot$ is the solar mass and the Sun is assumed to lie in the plane of the Galaxy. 
The best-fit normalisation for the \cite{Hunter24} and \cite{McMillan17} models are $(17.5\pm0.3)\times10^{-5}$ and $(18.1\pm0.3)\times10^{-5}$ (M$_\odot$ pc$^{-2}$ sr$^{-1}$) / (cm$^{-6}$ pc), respectively. 
We take the differences in the two profiles as an estimate of the systematic uncertainty on our knowledge of the stellar mass surface density of the Milky Way. 
These models match the emission measure well, indicating that the 0.7\,keV emission is primarily linked to the stellar mass distribution.

The bottom panel of Fig. \ref{fig:Coro} shows the binned emission measure of the 0.7\,keV component across four consecutive sky tiles. 
The profile is asymmetric, with the southern sky showing 10--15\% higher emission within $|b|\sim30^\circ$ of the Galactic disc. 
Flipping the latitude sign (grey points) highlights this asymmetry, with the southern hemisphere consistently brighter for low Galactic latitudes. 
The red line represents the stellar mass distribution {from \cite{Hunter24}}, but significant residuals ($\chi^2=58.2$ for 42 degrees of freedom; dof) suggest an unaccounted-for physical factor.

One possible explanation for the north-south asymmetry is the Sun’s position, which is approximately 10--25 pc above the Galactic plane \citep{Bland-Hawthorn16,Griv21}. 
To test this, we fitted the emission measures with the \cite{Hunter24} mass distribution, allowing the Sun’s height from the Milky Way mid-plane to be a free parameter in the fit (dashed red line in Fig. \ref{fig:Coro}). 
This significantly improves the fit ($\Delta\chi^2=-14.7$ from adding a free parameter), but the best-fit height ($h_\odot=47\pm14$~pc) is higher than expected. 
This is due to the asymmetry of the local stellar distribution, which is not included in our stellar mass model, as is further discussed in Section~\ref{sec:Gould}.
\begin{figure}[t]
\centering
\includegraphics[width=0.5\textwidth]{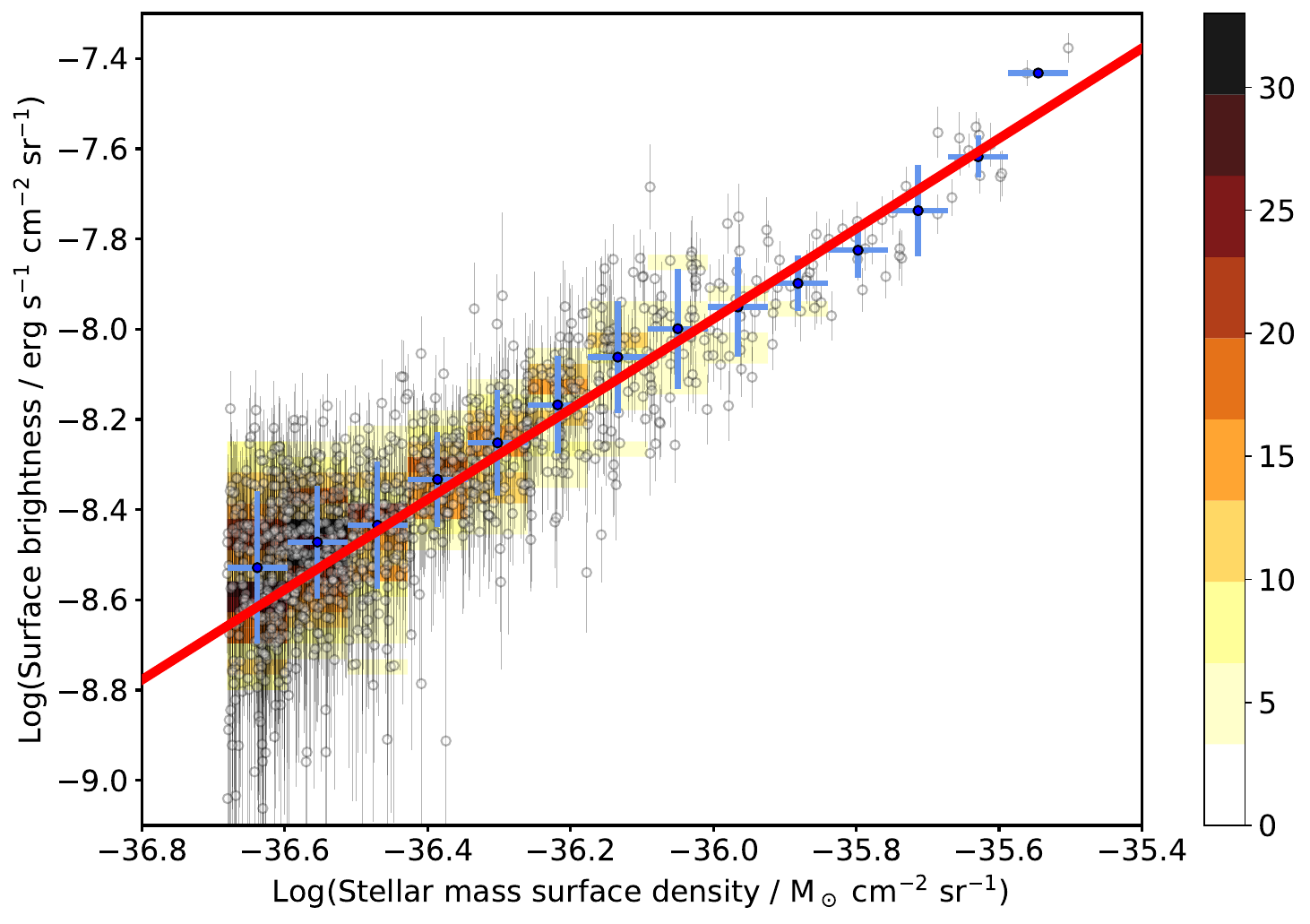}
\caption{Correlation between mass distribution of the Milky Way \citep{Hunter24} and total surface brightness, in the 0.1--10~keV band, of the 0.7~keV component within the sky area considered in this work (one data point for each sky tile). 
The surface brightness was computed extrapolating the best fit 0.7~keV component to the 0.1--10~keV band, and includes all emission, i.e. point sources and diffuse emission.
We rebinned the points into N bins along the $x$-axis, evenly spaced in $\log_{10,}$ of the stellar mass surface density. 
In each bin, the dark blue dot shows the average ($\log_{10}$) surface brightness, while the $x$ and $y$ error bars show the bin's half-width and the standard dispersion of the ($\log_{10}$) surface brightness in the bin, respectively. 
Each bin is also divided into other M sub-bins along the $y$-axis, with each sub-bin coloured according to the number of encompassed data points, as labelled in the colour bar on the right.
The red line shows the relation between surface brightness and stellar mass surface density with $L_x/M = 10.5\times10^{27}$ erg s$^{-1}$ M$^{-1}_\odot$, as derived from the fit  (Table \ref{Tab}).
}
\label{fig:CorrelHotMWPot}
\end{figure}

\section{Global correlation between 0.7~keV emission and stellar mass distribution}
\label{Sec:correl}

The latitudinal profile of the 0.7~keV emission closely follows the stellar mass surface density of the Milky Way. 
To confirm this across the entire western Galactic hemisphere, we fitted the surface brightness of the 0.7\,keV component in each sky tile using the mass-distribution model from \cite{Hunter24}, after applying the selection described in Sect.~\ref{sec:mask}, allowing the normalisation and the Sun’s height above the mid-plane to vary. 
The model provided a best fit of $\chi^2/{\rm dof}=1.34$, (see Table 1) with a height of the Sun above the plane ($h_\odot=27\pm5$~pc) consistent with independent measurements \citep{Bland-Hawthorn16,Griv21}. Strictly speaking, this fit is statistically unacceptable; however, as discussed in Section 4, most of the residuals arise from shot noise associated with resolved sources and from the asymmetric distribution of nearby stars, effects that are not captured by the stellar mass surface density model adopted here. 

\begin{table}
\footnotesize
\centering
    \caption{Best-fit parameters of the correlation between surface brightness of the 0.7~keV emission and the stellar mass surface density of the Milky Way. }
    \label{Tab}
    \begin{tabular}{c c c c c c c c c c c c c }
    \hline \hline 
\multicolumn{5}{c}{All} \rule{0pt}{3ex}   \\
           & {\bf M17}            & {\bf M17+$\beta$}   &  {\bf H24}        & {\bf H24+$\beta$}   \\
    \hline
EM         & \multicolumn{4}{c}{$7.7\times10^{-4}$} \\
Lx/M       & $11.8\pm0.1$   & $10.4\pm0.2$  & $10.5\pm0.1$ & $9.8\pm0.2$   \\
h$_\odot$  & $29\pm3$       & $23\pm4$      & $27\pm5$     & $23\pm5$      \\
n$_e$      & \ldots               & $0.3\pm0.1$   &  \ldots            & $0.2\pm0.1$   \\
$\chi^2$   & 1387.3         & 1310.9        & 1530.4       & 1510.5        \\
dof        & 1145           & 1144          & 1145         & 1144          \\
$p$        & $10^{-6}$ &$4\times10^{-4}$& $10^{-13}$   & $10^{-12}$\\
    \hline \hline
\multicolumn{5}{c}{Bright point sources removed} \rule{0pt}{3ex}  \\
           & {\bf M17}            & {\bf M17+$\beta$}   &  {\bf H24}        & {\bf H24+$\beta$}   \\
    \hline
EM         & \multicolumn{4}{c}{$5.2\times10^{-4}$} \\
Lx/M       & $7.62\pm0.07$ & $7.09\pm0.15$ & $6.87\pm0.06$ & $6.9\pm0.1$   \\
h$_\odot$  & $31\pm4$      & $28\pm4$      & $33\pm5$      & $34\pm6$      \\
n$_e$      & \ldots        & $0.17\pm0.09$ & \ldots        & $<0.1$        \\
$\chi^2$   & 1313.9        & 1295.9        & 1327.1        & 1326.6        \\
dof        & 1196          & 1195          & 1196          & 1195          \\
$p$        & $0.01$        & $0.02$        & $5\times10^{-3}$ & $4\times10^{-3}$\\
    \hline \hline
    \end{tabular}
\tablefoot{The surface brightness (SB) is in units of erg s$^{-1}$ cm$^{-2}$ sr$^{-1}$, the stellar mass surface density of the Milky Way is in units of M$_\odot$ cm$^{-2}$ sr$^{-1}$, following both the \cite{McMillan17} (M17), the \cite{McMillan17} plus a beta model component (M17+$\beta$), and the \cite{Hunter24} (H24) models. 
The luminosity per solar mass (L$_{\rm x}$/M) is in units of 10$^{27}$ erg s$^{-1}$ M$^{-1}_\odot$; the height of the Sun above the Galactic plane (h$_\odot$) is in parsecs; the density of the hot halo fitted with a beta model with $\beta=0.4$ and core radius of $R_c=1$~kpc is given in units of 10$^{-3}$ cm$^{-3}$ and it is the best-fit value at 10~kpc from the centre; then, the fit statistics ($\chi^2$) and degrees of freedom (dof) are given.}
\end{table}

Figure~\ref{fig:CorrelHotMWPot} demonstrates a strong correlation between the surface brightness of the 0.7\,keV component and the stellar mass surface density, with Spearman and Kendall’s coefficients being very high ($\rho = 0.774$ and $\tau=0.594$) and negligible p values ($p_0 = 10^{-230}$ and $p_0=10^{-200}$). 
This correlation suggests a significant physical connection (see Appendix \ref{sec:correl}) that is possibly associated with low-mass stars \citep{Guedel04,Masui09,Wulf19,Magaudda22,Caramazza23,Ampuku24}. 

If the 0.7~keV component were produced entirely by stars, the observed flux would be equivalent to the integral of the density of stars times their average 0.7\,keV luminosity. 
Assuming a universal initial mass function \citep{Kirkpatrick24}, M dwarfs with masses from 0.10--0.57~M$_{\odot}$ (M0--M6) compose about 68\% of the stars in the Milky Way and contribute to about 34\% of its stellar mass, while FGK stars contribute 39\% of the total stellar mass. 

From the best-fit linear relationship between mass distribution and surface brightness (red line in Fig. \ref{fig:CorrelHotMWPot}), we derived the X-ray luminosity per solar mass required to explain the observed 0.7\,keV emission of 
$L_{\rm x}/M_\odot=(1.05\pm0.01)\times10^{28}$ and $(1.18\pm0.01)\times10^{28}$ erg s$^{-1}$ M$^{-1}_\odot$ assuming the \cite{Hunter24} and \cite{McMillan17} descriptions of the mass distributions, respectively (Table \ref{Tab}).
To further test if low-mass stars could be responsible for the 0.7~keV emission, we made use of the X-ray brightness measured with \erosita\ for the F, G, K, and M dwarfs from the volume-complete sample of stars within $10$\,pc of the Sun published by \cite{Reyle21}. Under the assumption that the X-ray luminosity of the $10$\,pc stars is representative for the coronal emission of stellar X-ray sources in the large-scale galactic distribution, we can derive an independent measure of the mean $L_{\rm x}$ per stellar mass in the Galaxy from the X-ray properties of this sample.

The 10\,pc sample holds $103$ stars with spectral types\footnote{The spectral type is obtained from {\it Gaia} $G_{\rm BP} - G_{\rm RP}$ colours using the table maintained by E. Mamajek: https://github.com/emamajek/SpectralType.} between M0 and M6 and $30$ FGK-type stars in the western half of the Galactic sky. 
Zheng et al. (in prep) generated the mean stacked \erosita\ spectra for these two samples, one for the X-ray emission of all M stars and one for the FGK stars (see also Appendix \ref{sec:spec}). 
The two stacked spectra were fitted separately with a model composed of a few thermal components for an optically thin, collisionally ionised plasma ({\sc vapec} model in {\sc xspec}). 
The temperatures of the two hotter thermal components were assumed to be $kT = 0.7$ keV and $kT = 0.2$ keV, respectively.
Calculating the $0.1-10$\,keV flux of the $0.7$\,keV component in the best-fit model, dividing it by the number of stars that define the spectrum and scaling it to a distance of $10$\,pc, we determined an average X-ray luminosity of the 0.7\,keV component per M dwarf star in the western hemisphere of the 10 pc sample of $L_{\rm x, M}  = 1.9 \cdot 10^{27}\,{\rm erg~s^{-1}}$ (likely increasing to $L_{\rm x, M}  = 5.4 \cdot 10^{27}\,{\rm erg~s^{-1}}$, once the eastern hemisphere, where the four most luminous M dwarfs of the 10~pc sample reside, is also considered) and an average value of $L_{\rm x,FGK} = 1.07 \cdot 10^{28}\,{\rm erg~s^{-1}}$ for the FGK stars (Appendix \ref{sec:spec}). 

The average mass of M dwarfs in the $10$\,pc sample is $M_{\rm M} = 0.304 M_\odot$ (using the spectral type to mass conversion from \cite{Pecaut13}), such that the mean X-ray luminosity per solar mass of M dwarfs is $L_{\rm x,M}/M_{\rm M} = 6.3 \cdot 10^{27}\,{\rm erg~s^{-1} M_\odot^{-1}}$ (or $L_{\rm x,M}/M_{\rm M} = 17.7 \cdot 10^{27}\,{\rm erg~s^{-1} M_\odot^{-1}}$ in both hemispheres). 
Considering that the M dwarfs contribute $\sim 34\%$ of the total stellar mass budget of the Galaxy, they are expected to contribute an X-ray luminosity per total mass of $L_{\rm x} /M_\odot = 2.1 \cdot 10^{27}\,{\rm erg~s^{-1} M_\odot^{-1}}$ (or $L_{\rm x} /M_\odot = 6.0 \cdot 10^{27}\,{\rm erg~s^{-1} M_\odot^{-1}}$). 

The average mass of the FGK stars in the $10$\,pc sample is $M_{\rm FGK} = 0.903\,{\rm M_\odot}$.
Therefore, the mean X-ray luminosity per solar mass of FGK stars is $L_{\rm x, FGK}/M_{\rm FGK} = 1.18 \cdot 10^{28}\,{\rm erg~s^{-1} M_\odot^{-1}}$. 
Since FGK stars contribute $\sim$39\% of the total stellar mass, they contribute with an X-ray luminosity of $L_{\rm x}/M_\odot = 4.6\cdot 10^{27}\,{\rm erg~s^{-1} M_\odot^{-1}}$. 

We conclude that M dwarfs account for $(2.1/10.5)\cdot 10^{27} $erg s$^{-1}$ M$_\odot^{-1} \sim 20$\% of the observed 0.7\,keV emission, while their contribution would be as large as $(6.0/10.5)\cdot 10^{27} $erg s$^{-1}$ M$_\odot^{-1} \sim 57$\% if we consider the stars in both hemispheres of the 10~pc sample. 
Additionally, FGK stars account for $(4.6/10.5)\cdot 10^{27} $erg s$^{-1}$ M$_\odot^{-1} \sim 44$\% of the observed 0.7\,keV emission. 
Therefore, assuming that the 10~pc sample is representative of the entire Galaxy, we estimate that low-mass stars contribute $64-100$~\% of the 0.7~keV emission. 
This supports the hypothesis that the 0.7 keV emission primarily originates from the coronal activity of low-mass stars.

\section{Bright point sources and the asymmetric stellar structures around the Sun}
\label{sec:Gould}
\label{sec:PS}

The analysis presented in Sect. 3 assumes that the X-ray properties of stars in the 10 pc sample are representative of the entire Galaxy and that the simplified Galactic models used by \cite{Hunter24} and \cite{McMillan17} also accurately describe the stellar distribution in the vicinity of the Sun. However, fast-rotating low-mass stars, such as young stellar objects and intermediate-mass pre-main-sequence stars, are more luminous than field stars \citep{Preibisch05}. 
These stars are possibly under-represented in the 10~pc sample and are concentrated in nearby star-forming regions, which are seen to form structures tilted by $25-30^{\circ}$ relatively to the Galactic plane \citep[Appendix \ref{sec:starFrac}]{Guillout98,Alves20,Zucker22}.

\begin{figure*}[th]
\centering
\includegraphics[width=1.00\textwidth]{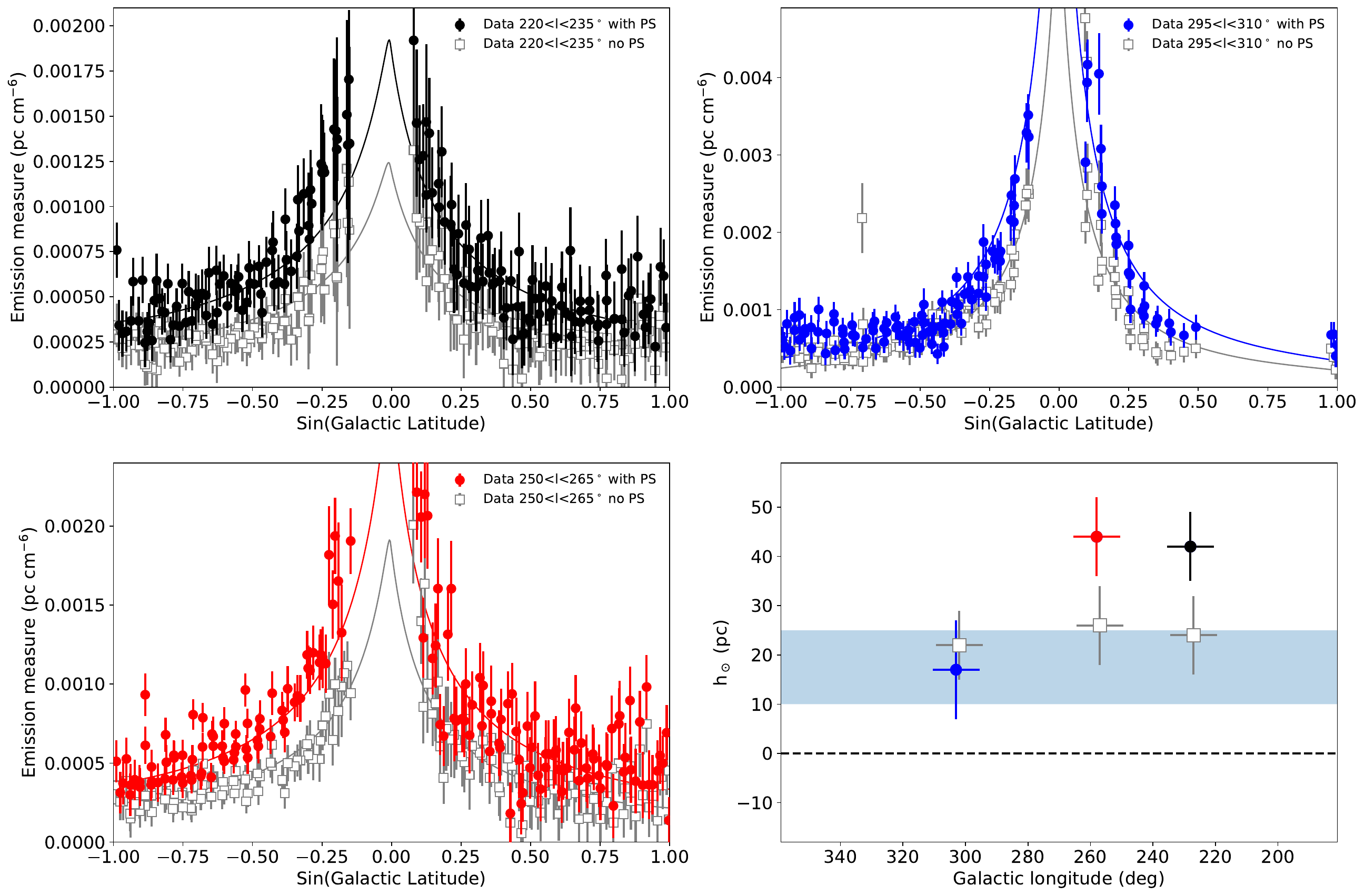}
\caption{Emission measures of 0.7\,keV component and its variations along three longitudinal stripes. 
The black, red, and blue data points show the emission measure of the 0.7\,keV component within the stripes: $220^\circ<l<235^\circ$; $250^\circ<l<265^\circ$; and $295^\circ<l<310^\circ$, respectively, which correspond to the same coloured dotted lines as in Fig. 1. 
The open grey squares show the emission measure of the 0.7\,keV component along the same stripes, once the bright point sources were removed. 
The coloured and grey lines show the best-fit stellar surface density model (with the \cite{Hunter24} model), which reproduces the emission measures where the bright point sources are either retained or removed, respectively. The fit was performed over the entire region considered in this work; it does not consider the addition of a beta model (see Table \ref{Tab}). 
The solid lines show the model's prediction (at the mean longitude of each stripe), which fits the entire region selected for our analysis. 
{\it Bottom right panel:} Best-fit position of the Sun above the Galactic plane ($h_\odot$) as a function of the Galactic longitude derived by fitting the 0.7\,keV emission measure along the same three longitudinal stripes. 
The filled coloured circles and the open grey squares show the best-fit offset of the Sun without and when removing the contribution from the bright point sources. 
The horizontal shaded blue interval shows the uncertainty on the offset of the Sun above the Galactic plane measured with other tracers \citep{Bland-Hawthorn16,Griv21}.} 
\label{fig:Multistripe}
\end{figure*}

The black, red, and blue data in the different panels of Fig. \ref{fig:Multistripe} show the emission measure of the 0.7~keV emission along longitudinal stripes within $220^\circ<l<235^\circ$, $250^\circ<l<265^\circ$ and $295^\circ<l<310^\circ$, respectively, as delineated on the sky map in Fig. 1, while the solid lines show the best fit with the \cite{Hunter24} model. 
Figure \ref{fig:Multistripe} shows that the emission measure is asymmetric not only along the stripe comprised
within longitudes of $220^\circ-235^\circ$ (see Fig. \ref{fig:Coro}), but
also along the stripe within $250^\circ-265^\circ$,  with the southern hemisphere appearing brighter\footnote{Figure \ref{fig:Multistripe} also shows that, on average, the emission measures at $250^\circ<l<265^\circ$ and $295^\circ-310^\circ$ are systematically larger, compared with the one at $220^\circ<l<235^\circ$. 
This can be easily interpreted considering the fact that these data are taken closer to the Galactic centre, where the stellar mass surface density is larger. }. 
However, the top right panel of Fig. \ref{fig:Multistripe} shows that the asymmetry is smaller at $295^\circ<l<310^\circ$. 
By fitting the observed asymmetry with an offset of the Sun above the Galactic plane, we observe values as large as $h_\odot\sim42-47$~pc for the black and red stripes. In contrast, the offset decreases to a value of $h_\odot=17\pm11$~pc within $295^\circ<l<310^\circ$ (see bottom right panel of Fig. \ref{fig:Multistripe}). 
This apparent dependence of the Sun's position on Galactic longitude is caused by the influence of nearby (young) stellar structures (Appendix \ref{sec:starFrac}) on the observed X-ray emission measure, which were not considered in our model \citep{Hunter24,McMillan17}. 

To further test whether the asymmetry observed in the longitudinal profile of the 0.7~keV emission is due to the asymmetric distribution of nearby luminous stars, we removed the point sources with count rates over $0.01$~ph~s$^{-1}$ in the 0.2--5~keV band from the \erosita\ data. 
This count-rate threshold roughly corresponds to sources brighter than $F_x>3.5\times10^{-14}$~erg~s$^{-1}$~cm$^{-2}$ for a spectral shape typical of low mass stars. Considering the variations within the \erosita\ all-sky exposure map (see \cite{Zheng24,Merloni24}), this threshold was selected to allow a homogeneous removal of point sources throughout the sky, therefore uniformly removing a portion of the emission from the most luminous stars around the Sun. 
In fact, stars with X-ray luminosities of $L_{\rm x}>10^{29}$ erg s$^{-1}$ are expected to be detected with \erosita, if located within 200 pc of the Sun (see Fig. \ref{fig:StarFrac}). 
After this adjustment, the best-fit offsets of the Sun along each of the three stripes at different longitudes considered in this work become consistent with expected values (see bottom right panel of Fig. \ref{fig:Multistripe}), confirming that the mismatch with our simplistic Galactic stellar mass model from Sect. \ref{sec:analysis} is primarily due to the asymmetric stellar distribution around the Sun. 
Indeed, the asymmetry in the 0.7~keV surface brightness closely follows the excess of bright stars within the solar neighbourhood (see Figs. \ref{fig:MapCoro}, \ref{fig:Multistripe} and \ref{fig:StarFrac}).

We then reassessed the global correlation between stellar mass and the residual 0.7~keV emission over the masked western Galactic hemisphere. 
Removing bright point sources improved the fit, $\Delta\chi^2=-203.3$ (see model H24 in Table \ref{Tab}), despite the higher number of degrees of freedom ($\Delta {\rm dof}=51$; i.e. the larger number of degrees of freedom is due to a larger number of sky tiles remaining after applying the filters described in Appendix \ref{sec:mask}\footnote{We find that, strictly speaking, this fit is also only marginally acceptable (at the $\sim99.5$ \% level). However, the removal of bright sources leads to a significant improvement in the fit, supporting the interpretation that the previously unacceptable fits were driven by shot noise from resolved point sources and by the asymmetric distribution of nearby stars. Although the brightest sources were removed, our best-fit results are likely still partially affected by these effects, albeit at a reduced level. }). 
With bright point sources removed, the mean surface brightness drops by about 33\%, indicating that resolved point sources contribute over a third of the 0.7\,keV emission. 

In conclusion, at least 33\% of the 0.7\,keV emission is from bright point sources. 
We estimate that low-mass stars contribute about 64-100\% of the 0.7\,keV component. However, this fraction may vary if the 10 pc sample does not fully represent the average Galactic stellar population. The remaining flux could potentially arise from younger, more X-ray-luminous stars or other star types. 

Contributions from hot diffuse plasma (e.g. hot interstellar medium, super-virial CGM) might also be present.
For example, a putative super-virial CGM component could be associated with outflows originating from the disc of the Milky Way, exhibiting a surface brightness that increases towards the disc and roughly follows the stellar mass surface density \citep{Stern24,Roy25}. 
However, we consider it unlikely that --if this were the case-- the implied luminosity per unit mass of the super-virial plasma would be consistent with the value observed in nearby stars, although we do not exclude that a super-virial CGM component might contribute to the observed surface brightness on some level. 
Indeed, the super-virial CGM component could be located at greater distances and primarily influenced by the dark-matter halo, in which case its distribution might follow a beta model, as discussed in the next section. 

\section{Possibility of a super-virial CGM contributing to the observed emission}

To test whether the data require a super-virial CGM component with a temperature of $kT\sim0.7$ keV, we added an emission component to the fit of the surface brightness of the 0.7~keV component, over the western Galactic hemisphere.
We postulate that the density of the super-virial plasma in the Milky Way halo follows a beta model, and its emission measure thus follows a squared beta profile. 
Since the core radius is unconstrained by the data, we assumed a radius of 1 kpc and report the plasma density at 10 kpc.

Adding super-virial CGM emission (0.7~keV) significantly improves the fit for the \cite{McMillan17} profile with $\Delta\chi^2=-76.4$ and $\Delta\chi^2=-18.0$ when point sources are retained or removed, respectively (models M17+$\beta$ in Table \ref{Tab}).
However, the improvement is marginal for the \cite{Hunter24} profile with the point sources removed. 
This suggests that detecting a super-virial CGM component with a beta-like profile is marginal and depends on uncertainties in the mass distribution profile.

The best-fit normalizations of the beta model (Table \ref{Tab}) vary, giving plasma densities at 10 kpc from $n_e=(3\pm1)\times10^{-4}$ cm$^{-3}$ to $n_e<1\times10^{-4}$ cm$^{-3}$, depending on the treatment of the stellar mass distribution and point-source subtraction. 
The statistical uncertainties are smaller than the systematic ones, so we conservatively set an upper limit on the super-virial CGM component at $n_e<4\times10^{-4}$ cm$^{-3}$ at 10 kpc. 
The right panel of Fig. \ref{fig:MapCoro} shows the emission measure of the 0.7~keV component after removing the contribution distributed as the stellar mass surface density. 
Clear patches of hot plasma emission are observed in the \erosita\ bubbles, Orion-Eridanus super-bubble, and other locations, supporting the presence of a hot interstellar medium in multiple regions of the Milky Way; moreover, at least in the case of the \erosita\ bubbles, the presence of the hot plasma is associated with an outflow \citep{Predehl20,Zhang24}.
Figure \ref{fig:MapCoro} shows excesses of hot plasma emission along some portion of the Galactic disc, in correspondence with known supernova remnants. 
Unfortunately, because of the large uncertainties on the profile of the mass distribution and on the correction for interstellar absorption affecting the 0.7~keV emission, it remains unclear whether a hot interstellar medium is present along the majority of the Galactic plane or not. Likewise, we leave the detailed comparison of absorption measurements and residual 0.7~keV emission observed here for future works.

\section{Conclusions} 

By studying the soft X-ray background in the western galactic hemisphere, we find that a dominant fraction of the 0.7\,keV emission, previously attributed to a super-virial CGM component, is in fact produced by low-mass stars. 
The remaining flux could be from younger or more active stars; in fact, at least 33\% of the 0.7\,keV emission is resolved into bright point sources. 
Only a small portion of the 0.7\,keV component flux may originate from a diffuse super-virial plasma. 
We also find hot super-virial plasma in localised regions (including the eROSITA bubbles), suggesting it is a truly diffuse component and that the super-virial plasma might be patchy and out-flowing \citep{Predehl20,Zhang24}. 

We stress that the average X-ray spectrum of low-mass stars exhibits multiple temperatures and can be approximately reproduced by two thermal components at $kT \sim 0.7$ and $\sim 0.2$~keV (Zheng et al. in prep). Therefore, low-mass stars may also contribute to the emission currently attributed to the virial CGM component, albeit to a much lesser extent.

The X-ray luminosity per stellar mass derived from the correlation and by stacking the 10~pc sample \citep{Zheng24} can help separate the contributions of low-mass stars from other sources in external galaxies, aiding in distinguishing emission from stars, the hot interstellar medium, the super-virial CGM, and the disc-halo interface.

\begin{acknowledgements}
The authors acknowledge Mattia Carlo Sormani, Barbara De Marco and Simone Scaringi for useful scientific discussion. We also thank the referees for the very useful comments, which greatly improved the clarity of the manuscript. 
GP acknowledges financial support from the European Research Council (ERC) under the European Union’s Horizon 2020 research and innovation program HotMilk (grant agreement No. 865637), support from Bando per il Finanziamento della Ricerca Fondamentale 2022 dell’Istituto Nazionale di Astrofisica (INAF): GO Large program and from the Framework per l’Attrazione e il Rafforzamento delle Eccellenze (FARE) per la ricerca in Italia (R20L5S39T9). 
EM is supported by Deutsche Forschungsgemeinschaft under grant STE 1068/8-1.
M.S. acknowledges support from the Deutsche Forschungsgemeinschaft through the grants SA 2131/13-1 and SA 2131/15-1.
MCHY and MF acknowledge support from the Deutsche Forschungsgemeinschaft through the grant FR 1691/2-1.

This work is based on data from eROSITA, the primary instrument aboard SRG, a joint Russian-German science mission supported by the Russian Space Agency (Roskosmos), in the interests of the Russian Academy of Sciences represented by its Space Research Institute (IKI), and the Deutsches Zentrum f\"ur Luft- und Raumfahrt (DLR). The SRG spacecraft was built by Lavochkin Association (NPOL) and its subcontractors, and is operated by NPOL with support from the Max Planck Institute for Extraterrestrial Physics (MPE).

The development and construction of the eROSITA X-ray instrument was led by MPE, with contributions from the Dr. Karl Remeis Observatory Bamberg \& ECAP (FAU Erlangen-Nuernberg), the University of Hamburg Observatory, the Leibniz Institute for Astrophysics Potsdam (AIP), and the Institute for Astronomy and Astrophysics of the University of T\"ubingen, with the support of DLR and the Max Planck Society. The Argelander Institute for Astronomy of the University of Bonn and the Ludwig Maximilians Universit\"at Munich also participated in the science preparation for eROSITA. 

The eROSITA data shown here were processed using the eSASS software system developed by the German eROSITA consortium.
\end{acknowledgements}

\bibliographystyle{aa} 
\bibliography{sn-bibliography}

\begin{appendix}
    
\section{Data analysis and spectral modelling}
\label{sec:ana}

This analysis used only photons from cameras with`on-chip' filters. Therefore we considered photons from the cameras: TM1, TM2, TM3, TM4, and TM6 \citep[see][for further details]{Predehl21}. For simplicity, we call this virtual camera TM8. To perform the analysis, we followed the regular grid of the sky tiles and produced a spectrum of TM8 for each $3^\circ\times3^\circ$ region. 
Instead of fitting the spectra of each camera separately, we have merged the spectra of all cameras for each sky tile (TM8), and we fit the merged spectra, using \texttt{pattern=15.} 
We assumed the abundances from \cite{Lodders03} and cross sections from \cite{Verner96}. 
We followed the analysis steps discussed in \cite{Yeung24}, including the removal of background flares, and we report here only the differences compared with that procedure. 
Differently from \cite{Yeung24}, all sources (point sources and clusters) are retained in the sky tile spectra. 
Although we acknowledge that removing detected point sources is advantageous for the study of diffuse emission, we chose to retain them to avoid biasing the estimates of the average luminosity per unit mass and the shape of the CXB.

In fact, the correlation between the 0.7~keV emission and stellar mass surface density can be used to derive the average luminosity per unit mass of the emitting sources, but only if all sources in the population are considered. 
If detected sources were removed from the total X-ray emission, we would need to subtract their contribution to the stellar mass surface density; otherwise, the derived luminosity per unit mass would be underestimated. 
However, accurately estimating the mass contribution of detected sources is nontrivial and prone to significant biases, depending on the correction method applied.
For this reason, we first retain point sources to determine the average luminosity per unit mass unbiasedly (see Sect. 2 and 3). 
We then compare these results with those obtained after excising detected sources (see Sect. 4). 
The other advantage of retaining point sources relies on the fact that the shape and intensity of the total emission associated with the CXB is well known, whereas further work is required to precisely determine how the CXB properties change when \erosita-detected sources above a given threshold are removed. 

The main drawback of this approach is that the presence of a single bright AGN or galaxy cluster with a flux exceeding $F_{2-10\,{\rm keV}} > 7 \times 10^{-11}$ erg cm$^{-2}$ s$^{-1}$ can roughly double the normalisation of the CXB component within a sky tile of about 9 square degrees. Fortunately, such bright sources are rare. We therefore excluded, a posteriori, all sky tiles affected by these very bright objects from our analysis, as detailed in Appendix D.

All spectral analysis was performed using the {\sc pyXspec} software package (version 2.1.0; \citep{Arnaud96}). 
We group each spectrum to have a minimum of 20 counts in each bin, and then we use the $\chi^2$ statistics to fit the CCD resolution spectra. 

We derived the ROSAT data points from the softest bands: R1 (which corresponds to ($E=0.11\text{--}0.284\,{\rm keV}$) and R2 ($E=0.14\text{--}0.284$\,keV).

We fit the spectra with the components described in \cite{Yeung24}, which are the emission from the un-absorbed local hot bubble ({\sc apec$_{\rm LHB}$}) plus the absorbed emission from the warm-hot component ({\sc apec$_{\rm WH}$}), the emission from the 0.7\,keV component ({\sc apec$_{\rm hGal}$}) and the emission from the cosmic X-ray background ({\sc bkn2pow}). 
In {\sc xspec} terminology this can be written as: {\sc apec$_{\rm LHB}$ +{disnht}(bkn2pow + apec$_{\rm WH}$ + apec$_{\rm hGal}$)}. 
The local hot bubble is assumed to have Solar abundances and temperature $kT=0.099$~keV \citep{Liu17}. 
The 0.7\,keV emission is assumed to have Solar abundances and temperature $kT=0.7$~keV.  
The warm-hot component is assumed to have an abundance of 0.1 Solar, and its temperature is a free parameter of the fit. 

We fitted the interstellar absorption with the {\sc disnht} model \citep{Locatelli22}, but we fixed the spread of column densities along the various lines of sight within the sky tile (we use a pixel size of $\sim12$~arcmin$^{2}$) to the value expected from the HI4PI map. 
We use the column density of neutral absorber estimated from the HI4PI survey \citep{HI4PIColl16} ($N_{\rm H,HI4}$, see Appendix \ref{sec:absorption}) as an initial guess for this parameter, which is free in the fit. 
We report the best fit column density as: $N_{\rm H,bf}$. 
All column densities are provided in cm$^{-2}$ units, and their logarithms are in decimal base, and we divide the column density by 1 cm$^{-2}$ to obtain a pure number. Throughout the paper, we express $\log (N_{\rm H}/{\rm 1 cm^{-2}})$ simply as $\log (N_{\rm H})$. 

We follow the definition of \cite{Ponti23} for the treatment of the CXB component, composed of a double broken power law. 
The CXB is assumed to have a photon index of $\Gamma=1.9$ below 0.4 keV, $\Gamma=1.6$ between 0.4 and 1.2 keV, and $\Gamma=1.45$ at higher energies. 
We do not explicitly model the combined emission from filaments of the warm hot intergalactic medium, galaxies, groups and clusters, because we expect it to be reproduced by our relatively soft broken power law CXB component. 

We model the instrumental background using the filter wheel closed (FWC) model\footnote{The model is available via \\https://erosita.mpe.mpg.de/dr1/AllSkySurveyData\_dr1/FWC\_dr1/.} presented in \cite{Yeung23}. 
Instead of fixing the normalisation of the FWC model, we start the fit of the spectrum of each sky tile in the 0.2-10 keV band, leaving the overall normalisation of the FWC model (as well as of the Al-K$\alpha$ line) to be a free parameter of the fit. 
In this way, the normalisation of the FWC component is anchored by the high-energy data, where the emission from the instrumental background dominates. 
Subsequently, we also fix the normalisation of all the FWC components and proceed with fitting the diffuse emission from the sky. 

Overall, there are six (${\rm EM_{LHB}}$, $\log{N_{{\rm H, bf}}}$, $kT_{\rm wh}$, ${\rm EM_{wh}}$, ${\rm EM_{hGal}}$, ${\rm norm_{CXB}}$) free parameters in the model.
We compute the error bars with the {\sc error} command corresponding to 90\% confidence level for one interesting parameter. 

Figure \ref{fig:spec} displays six spectra along the longitudinal stripe within $220^\circ<l<235^\circ$ and the best fit model. 
These spectra correspond to the green points shown in Fig. \ref{fig:Coro} and they are chosen to be representative of the major variations observed along this longitudinal stripe. 

\begin{figure}[t]

\includegraphics[width=0.45\textwidth]{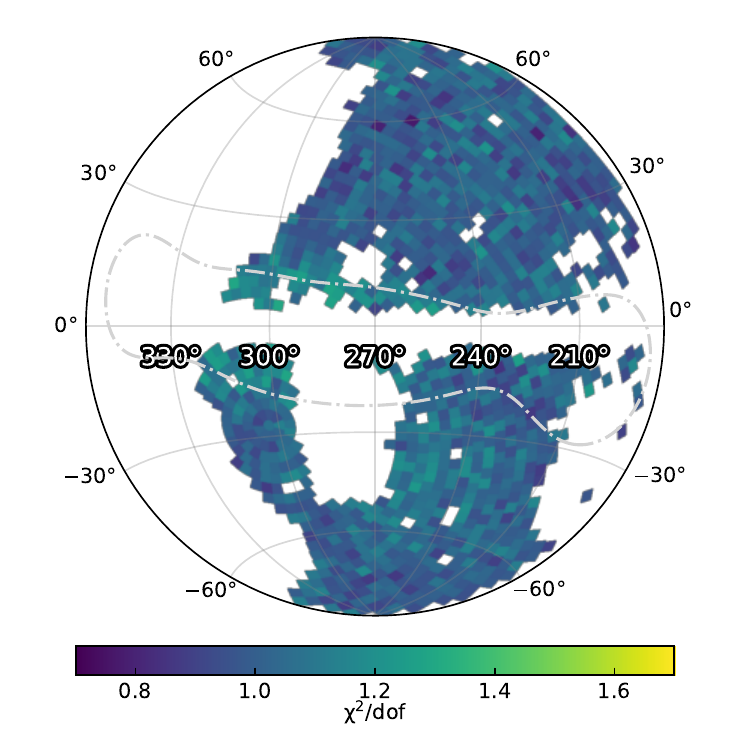}
\caption{Reduced $\chi^2$ of the spectral fit for each sky tile. The white areas show the sky tiles excluded from this study. Our simplistic model can reproduce the spectra from each sky tile in a statistically acceptable way. 
The dot-dashed grey contours indicate the region with the highest concentration of stars within 500~pc of the Sun.} 
\label{fig:MapChi2}
\end{figure}

Figure \ref{fig:MapChi2} demonstrates that our fiducial model provides a good description of the spectra observed by \erosita\ within all sky tiles studied in this work. 

In summary, our model reproduces the emission from the LHB, the CXB, the 0.7\,keV component investigated in this paper, as well as the warm-hot plasma with best fit temperature close to the virial temperature, therefore likely representing the hot plasma in the hot halo of the Milky Way.
Since the eRASS1 survey occurred during solar minimum, we neglect the solar wind charge exchange component as it was negligible \citep{Ponti23,Yeung23,Yeung24}.
The large reduced $\chi^2$ observed towards several sky tiles (Fig. \ref{fig:MapChi2}) indicates that our model would need to incorporate additional components at those locations. 
\section{Absorption from neutral (or low ionisation) material}
\label{sec:absorption}

One key aspect of this work is treating the complex effects of absorption on the measured spectra, which we address in more detail in this section. 

Following \cite{Locatelli24a}, we have estimated the total content of hydrogen in both atomic and molecular form $N_{\rm H}$ starting from the neutral atomic hydrogen column densities $N_{\rm H, HI4}$ from the full sky HI4PI radio survey \citep{HI4PIColl16} (see \cite{Locatelli24a}, Appendix A). 
We employed the same conversion as used in \cite{Locatelli24a}.  

\subsection*{Different column densities along various lines of sight: {\sc disnht}}
\label{sec:disnht}

By inspecting the HI4PI data, we observe that significantly different column densities of total hydrogen characterize the various line of sights within a sky tile. 
We attempted to reproduce such an effect by using the {\sc disnht} model, assuming a spread in column density as observed in HI4PI \citep{HI4PIColl16}. 

\subsection*{Single layer of absorption}
\label{sec:complexabsorption}

Our model assumes that the 0.7~keV component is fully absorbed by a single absorbing layer. 
However, in reality, both sources and absorbers are likely distributed along the line of sight, creating multiple layers that progressively attenuate more distant sources. 
This effect is expected to increase for larger column densities of absorption, therefore we selected only sky tiles with $\log(N_{{\rm H, HI4}})<21.5$. 

\begin{figure}[t]
\centering
\includegraphics[width=0.49\textwidth]{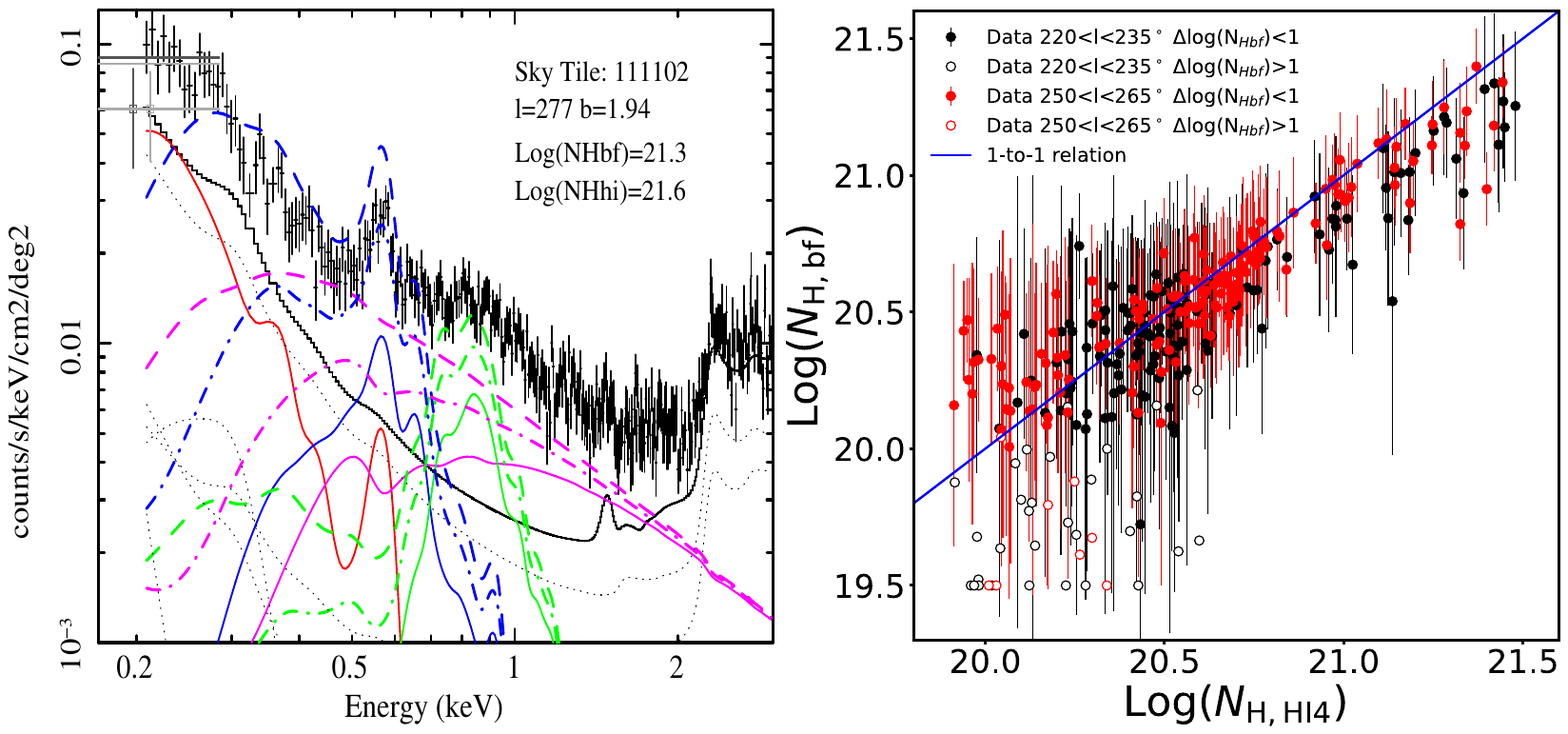}
\vspace{-2.8cm}

\includegraphics[width=0.49\textwidth]{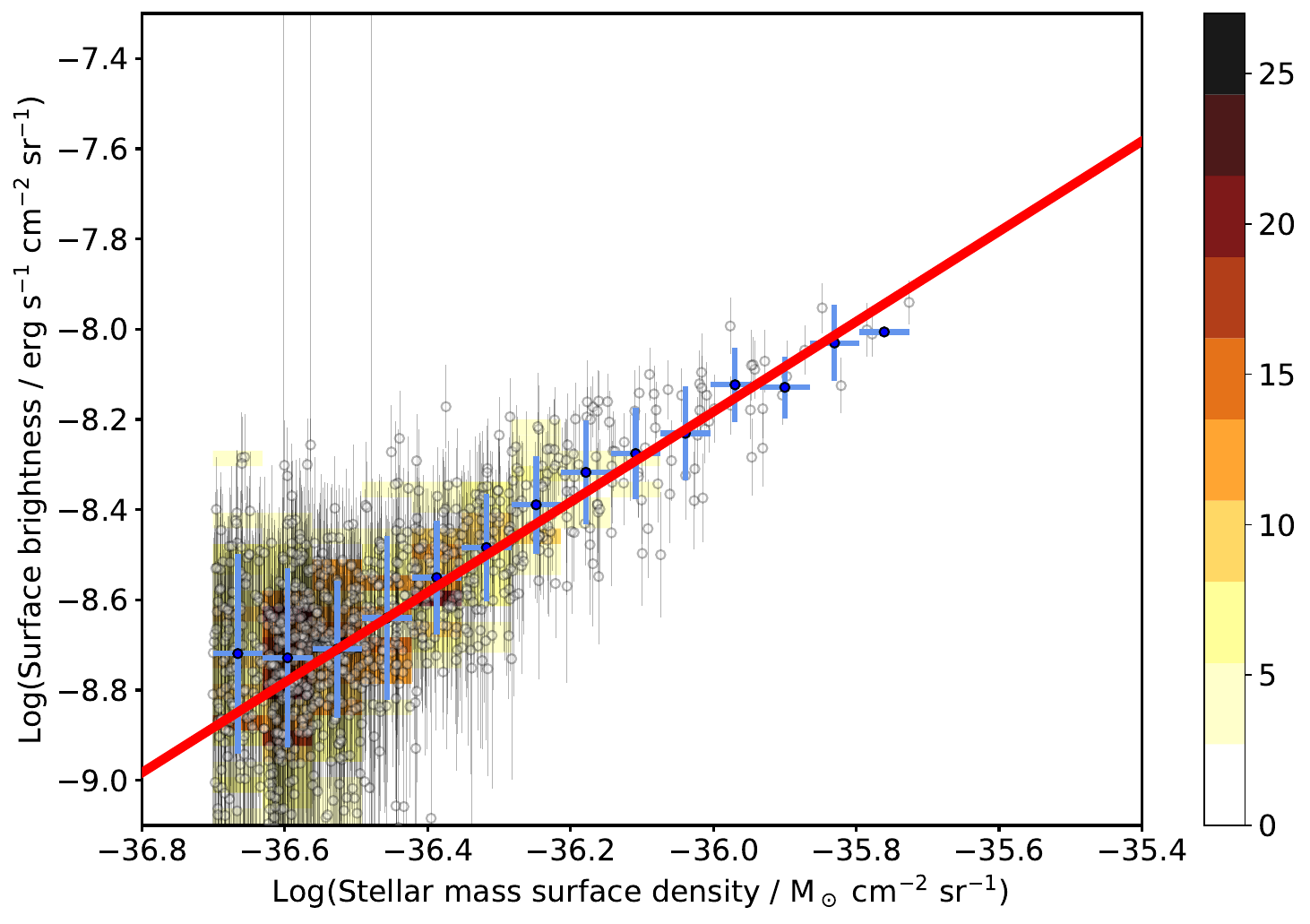}
\vspace{-0.4cm}

\caption{{\it Top left panel:} Best fit spectrum of the sky tile 111102, showing the effect of different column densities of absorbing material on the sky-background model. 
Black and grey data points show the \erosita\ spectrum and the \rosat\ constraints, respectively. 
The solid lines show the various components of the best-fit model with the FWC model in black, the unabsorbed local hot bubble in red, the warm-hot in blue, the 0.7 keV in green, and the cosmic X-ray background in magenta. 
This sky tile (shown in the bottom left of Fig. \ref{fig:spec} and in green in Fig. \ref{fig:Coro}) possesses a large total hydrogen column density ($\log(N_{{\rm H, HI4}})=21.6$), therefore it is not considered in this work. 
The best fit column density of absorbing material is $\log(N_{{\rm H, bf}})=21.3$. 
The dot-dashed and dashed lines show the same components (colour maintained) absorbed by a column density of $\log(N_{{\rm H, bf}})=21.0$ and $\log(N_{{\rm H, bf}})=20.5$, respectively. 
For column densities as large as $\log(N_{{\rm H, bf}})=21.0$ or $\log(N_{{\rm H, bf}})=21.3$, the presence of the low energy absorption cut off within 0.3-0.5~keV band induces a significant change on the shape and intensity of the warm-hot component. 
On the other hand, the shape of the 0.7\,keV component is almost unaffected up to $\log(N_{{\rm H, bf}})=21.3$ and its intensity drops by less than a factor of 2. 
{\it Top right panel:} Best fit column density of neutral absorption as a function of the total column density of Galactic absorption derived from the HI4PI survey for the single layer model. Black and red circles show the data within $220^\circ<l<235^\circ$ and $250^\circ<l<265^\circ$, respectively. Filled (and empty) circles indicate values which are constrained (and unconstrained; i.e., fit uncertainty larger 1\,dex) by the fit. The blue line shows the one-to-one correlation. 
{\it Bottom panel:} Correlation between the mass distribution of the Milky Way \citep{Hunter24} and the surface brightness as in Fig. \ref{fig:CorrelHotMWPot}, for sky tiles with $\log(N_{{\rm H, HI4}})<21.0$, once the bright point sources have been excised. The best fit correlation (red line) aligns with the previous results, confirming that absorption has a minimal effect. } 
\label{fig:SkyTile111102}
\end{figure}

To emphasise the role that absorption plays in the model, the top panel of Fig. \ref{fig:SkyTile111102} shows the best-fit spectrum of sky tile 111102. 
Although excluded due to its high column density ($\log(N_{{\rm H, HI4}})=21.6$), this sky tile serves as a clear example of absorption effects at high column densities. 
The best fit absorption column density ($\log(N_{{\rm H, bf}})=21.3$) is lower than the total hydrogen column density ($\log(N_{{\rm H, HI4}})=21.6$). 
This corroborates the idea that both emission and absorption are distributed along the line of sight; in other words, the total Galactic column density does not obscure the full emission column. 

The solid lines represent the best fit model components absorbed by the best fit column density of $\log(N_{{\rm H, bf}})=21.3$. 
To demonstrate the effect of absorption on individual spectral components, we show the warm-hot, the 0.7~keV component, and CXB models for two values of column densities, next to the best fit value. 
The dot-dashed and dashed lines in Fig. \ref{fig:SkyTile111102} show the emission from the warm-hot, the 0.7 keV and CXB components after being absorbed by a column density of $\log(N_{{\rm H, bf}})=21.0$ and $\log(N_{{\rm H, bf}})=20.5$, respectively. 
All other parameters, including temperature, abundance, and emission measure of each emission component, remain fixed at their best-fit value. 
Notably, for column densities as high as $\log(N_{{\rm H, bf}})=21.0$ or $\log(N_{{\rm H, bf}})=21.3$, the low-energy absorption cut off in the 0.3-0.5~keV band significantly alters both the shape and intensity of the warm-hot component. 
On the other hand, the shape of the 0.7 keV component is almost unaffected up to $\log(N_{{\rm H, bf}})=21.3$ and its intensity drops by about 20\% from $\log(N_{{\rm H, bf}})=20.5$ to $\log(N_{{\rm H, bf}})=21.0$ and by less than a factor of 2 from $\log(N_{{\rm H, bf}})=20.5$ to $\log(N_{{\rm H, bf}})=21.3$, therefore having a small effect on the results presented in this work. 

To further investigate the influence of absorption on our results, we have excluded all sky tiles with column density of absorption in excess of $\log(N_{\rm H})<21.0$. 
The bottom panel of Fig. \ref{fig:SkyTile111102} shows that, even with the more restrictive limit of $\log(N_{\rm H})<21.0$, we obtain a correlation ($L_{\rm x}/M=(6.7\pm0.01)\times10^{27}$~erg s$^{-1}$~M$_{\odot}^{-1}$), once the point sources are removed, consistent with the result reported in Sect. 4 and Table 1 (Model H24), therefore corroborating the main results of this work. 
We have repeated this exercise with point sources retained and obtain a correlation ($L_{\rm x}/M=(1.04\pm0.01)\times10^{28}$~erg s$^{-1}$~M$_{\odot}^{-1}$) consistent with the result reported in Sect. \ref{Sec:correl} and Table 1, further confirming that the main results of this work do not depend on the subtraction of the bright point sources. 

The top right panel of Fig. \ref{fig:SkyTile111102} shows the best fit column density of neutral absorption for the single layer model $\log(N_{{\rm H, bf}})$ versus the Galactic total column density along that line of sight $\log(N_{{\rm H, HI4}})$, for the two stripes within $220^\circ<l<235^\circ$ and $250^\circ<l<265^\circ$ in black and red, respectively. It is observed that the best-fit column density is typically smaller than the total Galactic value, as in sky tile 111102, discussed above in detail, therefore corroborating the inference that the emission and absorption are distributed along the line of sight. 
Figure \ref{fig:SkyTile111102} also shows that when the total Galactic column density is low, the best-fit column density of absorbing material ($\log(N_{{\rm H, bf}})$) might sometimes result in being pegged at its lowest value. 
We report the best-fit value in the top-right panel of Fig. \ref{fig:SkyTile111102} as an open circle when the uncertainties are larger than 1\,dex.

\subsection*{Multiple layers of absorption}
\label{sec:multilayer}

Considering that both emission and absorption are distributed along the line of sight—and that the warm-hot and 0.7~keV emissions likely have different spatial distributions—raises the question of whether a single-layer absorption model adequately represents this complexity. 
To test this, we refitted all sky tiles within the western Galactic hemisphere using the same spectral model, which includes unabsorbed (LHB) and absorbed (CXB, warm-hot, and 0.7~keV emission) components. 
However, this time, we allowed each absorbed component to have a different column density, referring to this approach as a multi-layer absorption model. 
Specifically, we assume that the CXB is absorbed by the total Galactic column density, while the absorption towards the warm-hot Galactic emission is left free ($19.5\leq\log(N_{{\rm H, bf, wh}})$), and the absorption towards the 0.7~keV emission is allowed to vary within ($19.5\leq\log(N_{{\rm H, bf, hGal}})\leq\log(N_{{\rm H, HI4}})$). 

\begin{figure}[t]
\hspace{-2cm}

\includegraphics[width=0.49\textwidth]{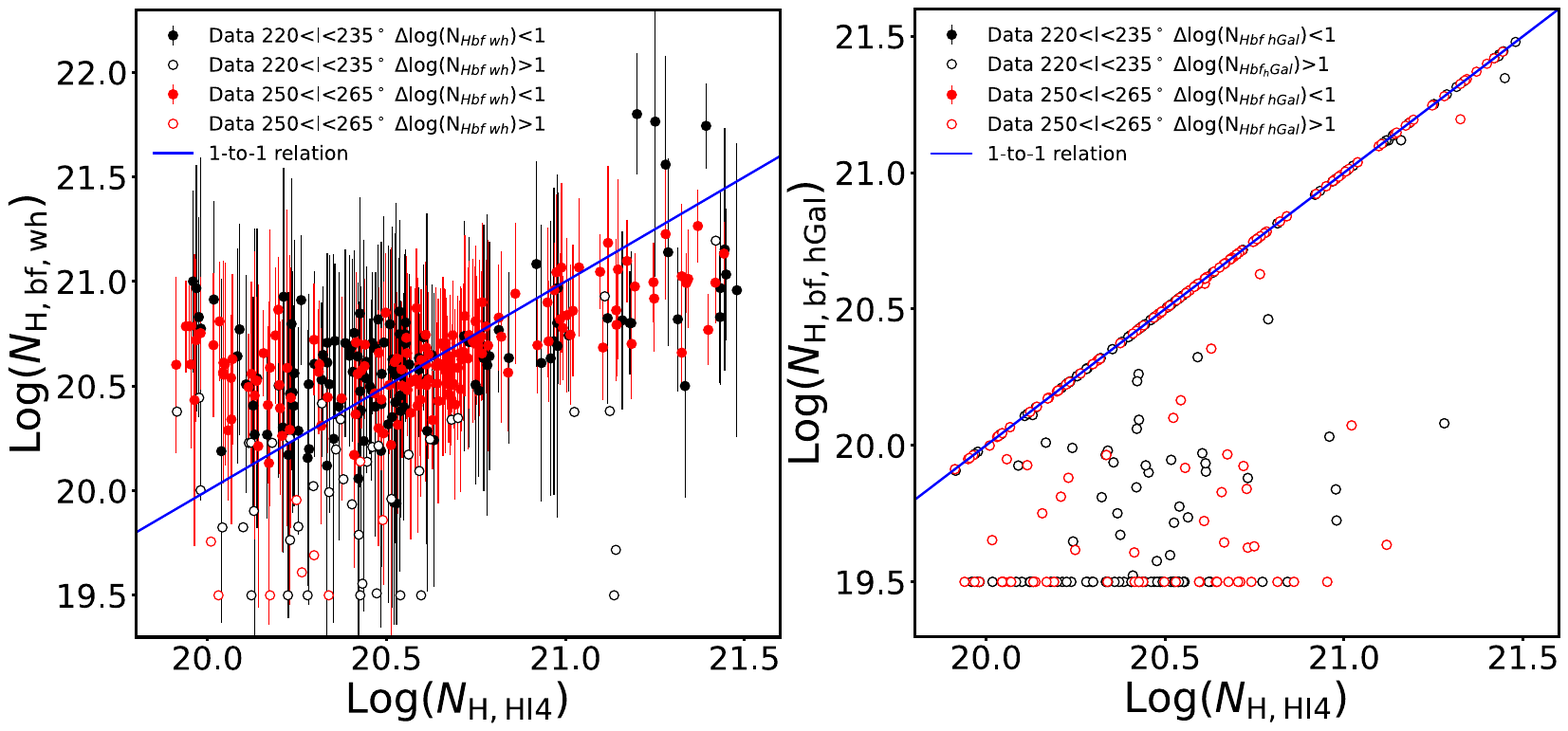}
\vspace{-2.8cm}

\includegraphics[width=0.5\textwidth]{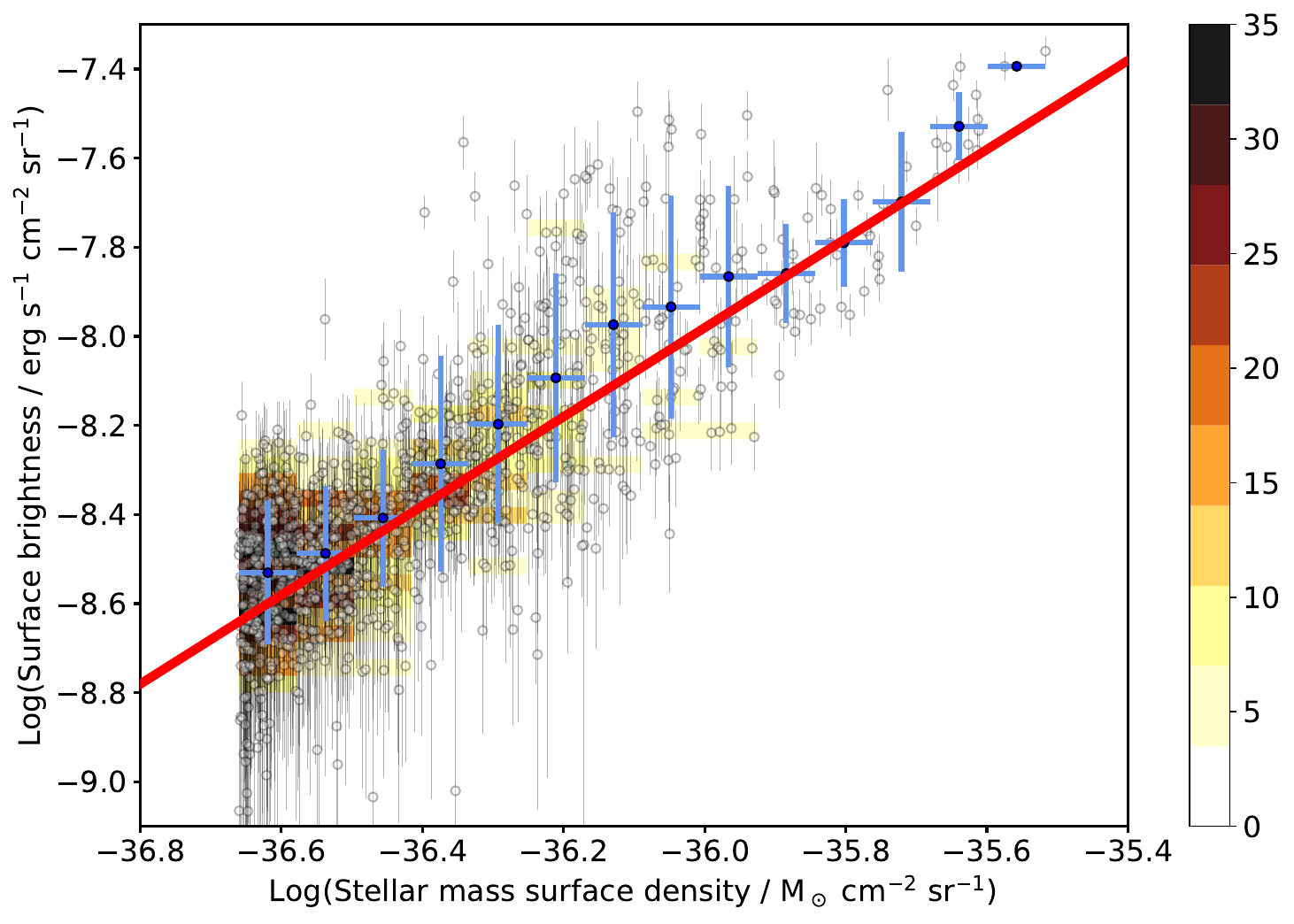}
\caption{Effects of different absorption models on the best-fit results.  
{\it Top-left and top-right panels:} Column density of neutral absorption of the warm-hot (left) and 0.7~keV (right) components in the multiple-layer model. Same colours and symbols as in the top-right panel of Fig. \ref{fig:SkyTile111102}.
The column density of the 0.7~keV component is unconstrained and often pegged at very low values or the total Galactic column density. 
{\it Bottom panel:} Correlation for the multilayer absorption model. 
Despite the unconstrained column density of the 0.7~keV component inducing a larger scatter in the best-fit surface brightness of the 0.7~keV emission, the correlation with the stellar mass surface density persists with a correlation coefficient ($L/M$) consistent with the single-layer model. 
}
\label{fig:multilayer}
\end{figure}

The top-left and top-right panels of Fig. \ref{fig:multilayer} show the best-fit column densities from this approach for the warm-hot and 0.7~keV components, respectively. 
The column density of the 0.7~keV component is poorly constrained, ranging from the lowest possible value ($\log(N_{{\rm H, bf, hGal}})=19.5$) to the maximum value allowed ($\log(N_{{\rm H, bf, hGal}})=\log(N_{{\rm H, HI4}})$; Fig. \ref{fig:multilayer} top right panel). 
Indeed, the uncertainties on $\log(N_{{\rm H, bf, hGal}})$ are large enough that the best fits are often pegged either at the maximum or minimum values allowed. 
This is expected, as the 0.7~keV component has a fixed spectral shape due to its set temperature and metal abundances. 
Moreover, as shown in Fig. \ref{fig:SkyTile111102} (top-left panel), column densities below $\log(N_{{\rm H, bf, hGal}})<21.5$ have little effect on the 0.7~keV component, leading to the fact that $\log(N_{{\rm H, bf, hGal}})$ is poorly constrained.
The top-left panel of Fig. \ref{fig:multilayer} indicates that the best-fit column density for the warm-hot Galactic component follows a similar trend to the single-layer model but with larger uncertainties due to the few constraints on $\log(N_{{\rm H, bf, hGal}})$.

The bottom panel of Fig. \ref{fig:multilayer} shows the correlation between the surface brightness of the 0.7~keV component and the stellar mass surface density in the multi-layer model. 
Despite the few constraints on $\log(N_{{\rm H, bf, hGal}}$), which introduces significant scatter, the correlation remains evident. 
The best-fit correlation yields $L_{\rm x}/M=(1.05\pm0.01) \times 10^{28}$ erg s$^{-1}$ M$_\odot^{-1}$, fully consistent with the single-layer model. 
This confirms that the relationship between the surface brightness of the 0.7~keV component and the stellar mass surface density remains largely unaffected by low column densities of obscuring material ($\log(N_{{\rm H, HI4}})<21.5$) as selected in this work.

\section{Definition of the emission measure}
\label{sec:EM}

The normalisation of the {\sc XSpec} model {\sc APEC} component is defined as: 
$N_{apec} = \frac{10^{-14}}{4\pi D^2} \int{n_e n_H dV}$, where the distance to the source $D$ is given in cm, $dV$ is the volume element in cm$^{3}$ and $n_e$ and $n_H$ are the electron and hydrogen densities in cm$^{-3}$, respectively. 
Therefore, the {\sc apec} normalisation has units of 10$^{-14}$/4$\pi$ cm$^{-5}$. 
If the {\sc apec} normalisation is multiplied by volume emissivity (with units of [erg s$^{-1}$ cm$^{3}$]), then the {\sc apec} normalisation becomes consistent with a flux (with units of [10$^{-14}$/4$\pi$ erg s$^{-1}$ cm$^{-2}$]). 

For extended sources, such as the diffuse emission studied in this work, the spectra we analyse do not cover the entire extent of the source; therefore, the {\sc apec} normalisation will be performed only on a fraction of the volume.  
We define as emission measure ${\rm EM} = \int{n_e n_H dl}$ with units [pc cm$^{-6}$], which can be derived from the normalisation of the {\sc apec} component: $N_{apec} [{\rm cm^{-5}}] = \frac{10^{-14}}{4\pi D^2} \int{n_e n_H dV} = \frac{10^{-14}}{4\pi D^2} 3.09\times10^{18}[{\rm cm/pc}] D^2 \Omega \int{n_e n_H dl}$ [pc cm$^{-6}$] $= 3.09\times10^{4} [{\rm cm/pc}] \frac{\Omega}{4\pi}$ EM [pc cm$^{-6}$], where $\Omega$ is the solid angle in steradian and $l$ is the depth, along the line of sight, of the emitting source.

\section{Creation of the mask}
\label{sec:mask}

To remove the contribution from bright sources that have not been modelled in our fits, we have filtered out the sky tiles that are affected by these additional emission components.
We selected the tiles to be removed based on the following criteria:  
\begin{enumerate}
\item{} the CXB component has a large normalisation (${\rm norm_{CXB}}>0.0035$ photons s$^{-1}$ cm$^{-2}$ keV$^{-1}$ deg$^{-2}$ at 1 keV), which is about 35~\% larger than the expected value; 
\item{} the column density of neutral absorption is in excess of $\log(N_{\rm H, HI4}$)$>$21.5; 
Indeed, column densities in excess of this value might affect the proper estimation of the emission measure of the 0.7~keV emission; 
\item{} the Galactic latitude is within $|b|<4^\circ$, therefore removing the emission from the Galactic plane and the sources therein; 
\item{} the centre of the tile lies within the footprint of the \erosita\ bubbles; 
\item{} the centre of the tile lies within the footprint of the Orion-Eridanus superbubble ($-60^\circ<b<0^\circ$ and $180^\circ<l<210^\circ$) and the emission measure of the warm-hot CGM component is large (EM$_{\rm wh}>0.25$ pc cm$^{-6}$); 
\item{} the centre of the tile lies within the footprint of the LMC and Goat's Horn ($-51^\circ<b<0^\circ$ and $260^\circ<l<294^\circ$) and the emission measure of the warm-hot CGM component is large (EM$_{wh}>0.25$ pc cm$^{-6}$); 
\item{} the centre of the tile lies within the footprint of the Gemini-Monoceros X-ray enhancement ($0^\circ<b<30^\circ$ and $180^\circ<l<210^\circ$) and the emission measure of the warm-hot CGM component is large (EM$_{\rm wh}>0.25$ pc cm$^{-6}$); 
\item{} the centre of the tile lies within the footprint of the Antlia supernova remnant ($12^\circ<b<22^\circ$ and $268^\circ<l<280^\circ$). 
\end{enumerate}

{\bf Bright X-ray binaries, galaxy clusters and AGN.} 
We note that emission from bright galaxy clusters and AGN are removed from criterion one (${\rm norm_{CXB}}>0.0035$ photons s$^{-1}$ cm$^{-2}$ keV$^{-2}$ at 1 keV), while bright X-ray binaries are removed from criteria one and three. 

{\bf Large column densities of neutral absorption, Galactic disc and sources therein.} 
At large column densities of neutral absorption, the emission measure of the 0.7~keV emission becomes less reliable; therefore, we filter that out (criterion 2). This selection removes most sky tiles with centres within a few degrees of the Galactic plane. However, we observe that most of the remaining sky tiles within $4^\circ$ and $-4^\circ$ from the plane are affected by nearby sources, which would require a more accurate treatment of their emission. For this reason, we do not consider the sky tiles with centre within $|b|<4^\circ$. 

{\bf Large extended features.} 
We first selected broad ranges in Galactic longitudes and latitudes which appear affected by the various soft X-ray emitting extended features, then we removed the sky tiles with emission measure of the CGM component in excess of EM$_{\rm wh}>0.25$ pc cm$^{-6}$. 

The final mask excludes 1071 sky tiles, corresponding to approximately 48~\% of the western sky. 
The majority of this area (1024 sky tiles, or 46~\%) is removed by the selection criteria outlined in points 3-8, aiming at removing the large extended features. 
By discarding sky tiles with $\log(N_{\rm H, HI4}$)$<$21.5, 24 sky tiles are excluded at Galactic latitudes larger than $|b|<4^\circ$ and 156 sky tiles are already removed along the Galactic plane by selection 3.
Finally, applying the criterion on the CXB component (${\rm norm_{CXB}}<0.0035$ photons s$^{-1}$ cm$^{-2}$ keV$^{-1}$ deg$^{-2}$ at 1 keV) eliminates a further 23 sky tiles, corresponding to about 1~\% of the western hemisphere. 

\begin{figure}[t]
\centering
\includegraphics[width=0.48\textwidth]{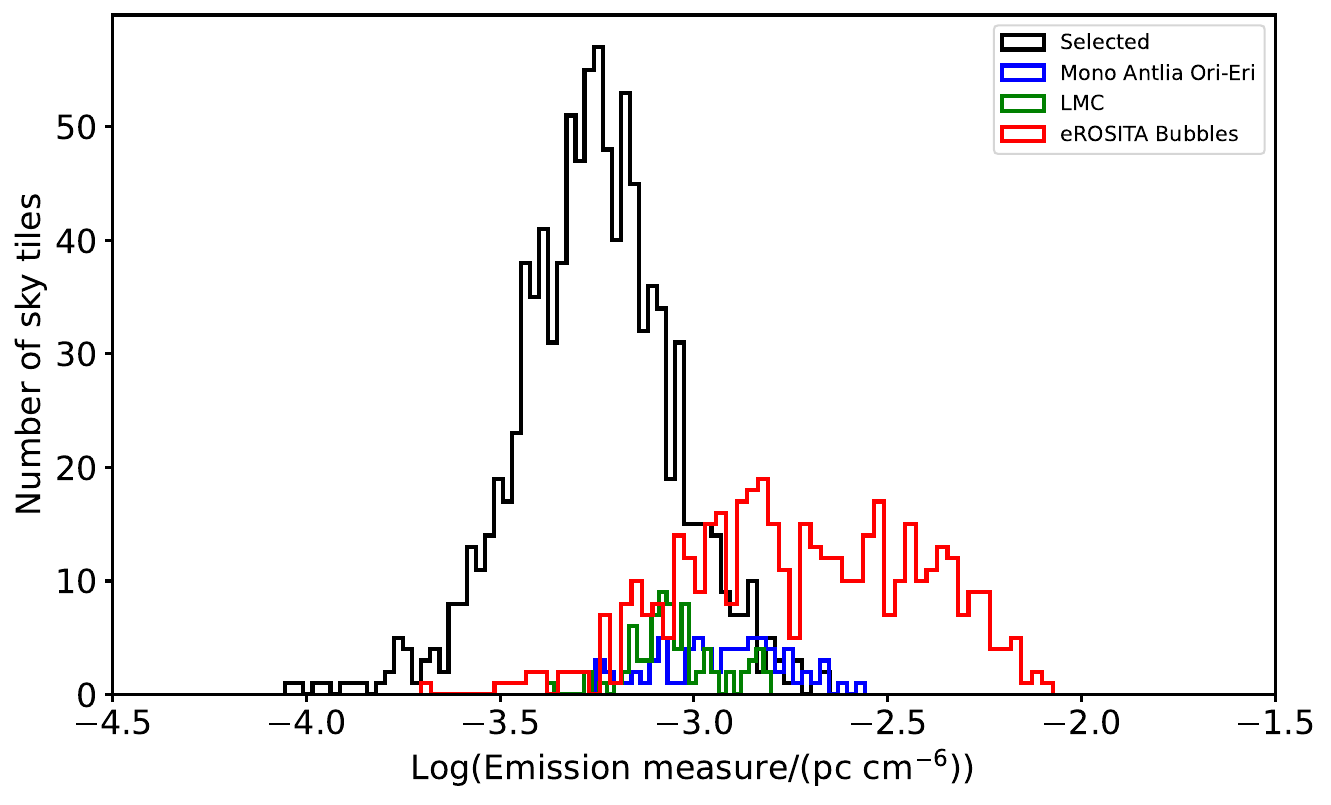}
\caption{Histograms of the emission measure of the 0.7~keV component, after applying selection criteria 1--3 (see Appendix~\ref{sec:mask}), for the sky tiles within the \erosita\ bubbles (red), the LMC and Goat's Horn (green), the Gemini-Monoceros X-ray enhancement, Antlia and the Orion-Eridanus superbubble (blue) and our final selection (black). }
\label{fig:Histo}
\end{figure}
The black, red, green and blue lines in Fig. \ref{fig:Histo} show the histograms of the emission measure of our final selection and for comparison, the histograms of the footprint of the \erosita\ bubbles, the LMC and Goat's Horn and the Gemini-Monoceros X-ray enhancement, Antlia and Orion-Eridanus superbubble, respectively, after applying the criteria from items one to three of the above list. 
The emission from these extended sources is not considered in our simplified model, which might falsely attribute it to the 0.7~keV component. Therefore we filtered out these extended features \citep{Ponti19,Predehl20,Gupta23,Zhang24,Locatelli24b,Knies24}.

\section{On the relation between X-ray surface brightness and stellar mass surface density}
\label{sec:correl}

In this section, we make the physical connection behind the correlation between the X-ray surface brightness and stellar mass surface density more explicit. 

What is the X-ray flux emitted by all stars within a shell C at distance $r$ and thickness $dr$?
We consider a model for the mass density distribution $\rho(r)$ of stars in the Milky Way \citep{Hunter24}.
The mass $M_C(r)$ contained in the shell at distance $r$ is:

\begin{equation}
    M_C(r) = \rho(r) \cdot 4\pi r^2 dr
    \label{eq:shell_mass}
\end{equation} 

Now, let us consider all stars in a galaxy. They produce a total X-ray luminosity $L_X$ regardless of their single properties. If the average properties of the population do not change across the galaxy, the luminosity of a given region $\tilde{\cdot}$ will be proportional to the number of stars included within the boundaries of $\tilde{\cdot}$, that is in turn proportional to the stellar mass of the region $\tilde{M}_*$. 
The constant of proportionality $\alpha$ between the luminosity $\tilde{L}_X$ and mass $\tilde{M}_*$ of the region is then an average luminosity per unit mass $\langle L_X / M_\odot \rangle$ so that

\begin{equation}
    \tilde{L}_X = \alpha \cdot \tilde{M}_* \,\,\,\,\, \rm with \,\,\,\,\, \alpha \equiv \left\langle \frac{L_X}{M_{\odot}} \right\rangle 
    \label{eq:region_L}
\end{equation} 

The X-ray luminosity of the shell $L_{X,C}(r)$ can then be obtained by combining the above equations

\begin{equation}
    L_{X,C}(r) = \alpha \cdot \rho(r) \cdot 4\pi r^2 dr
    \label{eq:shell_L}
\end{equation} 
By sitting at the geometrical centre of the 3D shell, an observer receives a flux $F_{X,C}$ from the shell, computed as

\begin{equation}
    F_{X,C}(r) \equiv \frac{L_{X,C}(r)}{4\pi r^2} = \alpha \cdot \rho(r) dr
    \label{eq:shell_flux}
\end{equation} 

The flux from the shell within an infinitesimal solid angle $d\Omega$ is thus
\begin{equation}
    \frac{dF_X}{d\Omega}(r) = \alpha \cdot \frac{\rho(r)}{4\pi} dr
    \label{shell_fluxPerAngle}
\end{equation}
so that 
\begin{equation}
    \int_{4\pi} \frac{dF_X}{d\Omega'}(r)\, d\Omega' = F_{X,C}(r)
\end{equation}
and in general
\begin{equation}
    F_X(r,\Omega) \equiv \int_\Omega \frac{dF_X}{d\Omega'}(r)\, d\Omega' = \frac{\Omega}{4\pi} \cdot \alpha \cdot \rho(r) dr
\end{equation}
subtended by a generic solid angle $\Omega$.

By integrating the latter equation along the line of sight, we obtain the formula to compute the expected X-ray flux within the solid angle $\Omega$ as a function of the stellar mass density model $\rho(r)$
\begin{equation}
    F_X(\Omega) = \frac{\Omega}{4\pi} \cdot \alpha \int_0^\infty \rho(r') dr' \,\,\,\,\,\,\,\rm  \left[ erg\, s^{-1} cm^{-2} \right]
    \label{eq:cone_int_flux}
\end{equation}

The measured X-ray surface brightness $S_X$ is then
\begin{equation}
    S_X \equiv \frac{F_X(\Omega)}{\Omega} = \alpha \int_0^\infty \frac{\rho(r')}{4\pi} dr' \, \, \, \, \, \, \, \, \, \,\rm \left[ erg\, s^{-1}\, cm^{-2}\, sr^{-1} \right]
    \label{eq:cone_S}
\end{equation}

We now define the surface mass density per unit solid angle $\Sigma_M$ as
\begin{equation}
    \Sigma_M \equiv \int_0^\infty \frac{\rho(r')}{4\pi} dr' \, \, \, \, \, \, \, \, \rm \left[M_\odot\, cm^{-2}\, sr^{-1}\right]
    \label{eq:sigma_M}
\end{equation}

Therefore, the surface brightness $S_X$ (eq.~\ref{eq:cone_S}) can be simply recast as
\begin{equation}
    S_X = \alpha \cdot \Sigma_M 
\end{equation}

so that
\begin{equation}
    \left\langle \frac{L_X}{M_{\odot}} \right\rangle = \frac{S_X}{\Sigma_M} \, \, \, \, \, \, \, \, \, \, \,\rm  \left[ erg\, s^{-1}\, M_\odot^{-1}\right]
\end{equation}
Therefore, the correlation coefficient we measure allows us to determine the $L_x/M$ of the sources producing the 0.7~keV emission. 

Clearly, the X-ray luminosity of stars can be very different for stars of different types, ages, etc. 
For example, the X-ray luminosities of the different stars of the M dwarf 10~pc sample span four orders of magnitude \citep{Caramazza23}. 
However, the average luminosity of the sample is expected to show less scatter. 
Accordingly, to determine the average luminosity per stellar mass of low mass stars, we have stacked the 0.7~keV emission over the entire volume of the complete 10~pc sample. 
Clearly, the fractional contribution of stars to the 0.7~keV emission is valid under the assumption that the 10~pc sample is representative of the average Galactic population. 

Studies of stars close to the Sun show that the various sub-classes of stars have different luminosity per mass, and their spatial distribution can be somewhat different from sub-class to sub-class \citep{Bovy17}.
However, if the luminosity per mass of stars were very different in various parts of the Galaxy, the correlation between stellar mass density and 0.7 keV emission would either not be present or possess a large scatter. This suggests that part of the small scatter in the correlation can be associated with the different spatial distributions of the various sub-classes of low mass stars, as indicated by studies of stars close to the Sun \citep{Bovy17}.

The excess of 0.7~keV emission associated with the nearby young stellar structures suggests that the 10~pc sample might be missing very luminous stars. 
If so, low mass stars might contribute even more than 70~\% of the 0.7~keV emission. 
Complete samples of low mass stars which extend to significantly larger volumes are needed to determine whether the 10~pc sample is representative of the average population.

\section{Mean spectra of the M dwarf and FGK stars}
\label{sec:spec}

\begin{figure*}[t]
\centering
\includegraphics[width=0.495\textwidth]{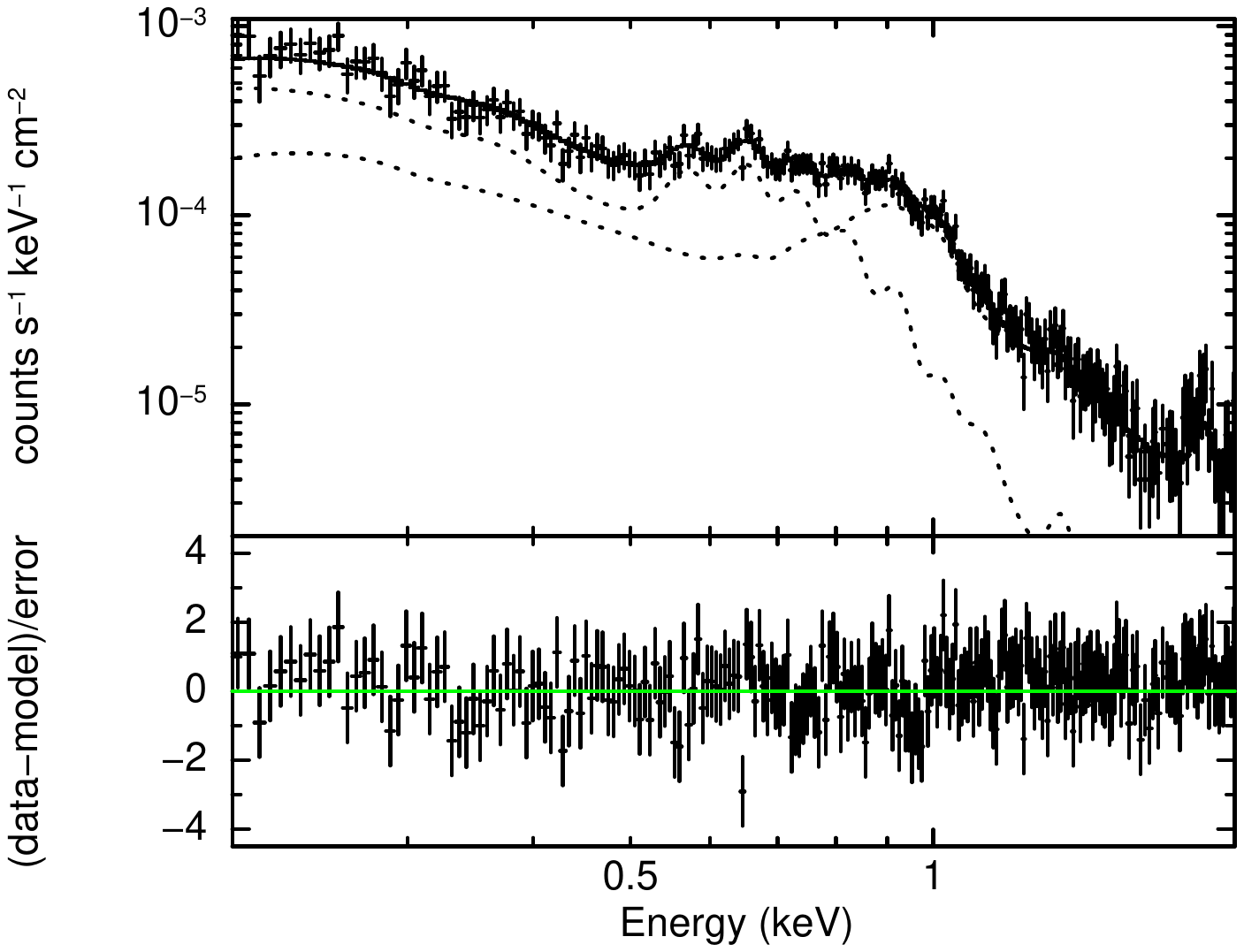}
\hspace{-1cm}
\includegraphics[width=0.495\textwidth]{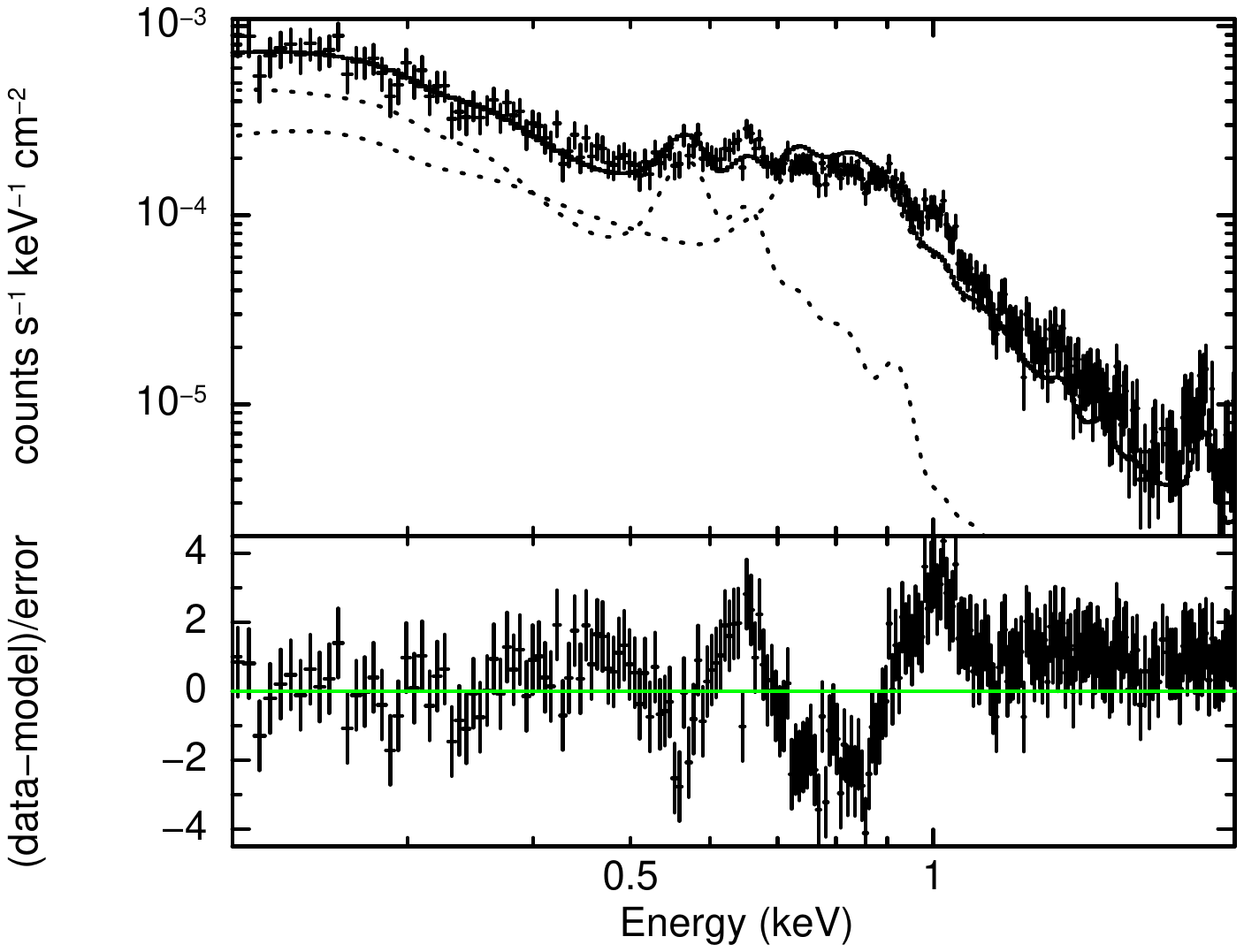}
\vspace{-1.1cm}

\includegraphics[width=0.495\textwidth]{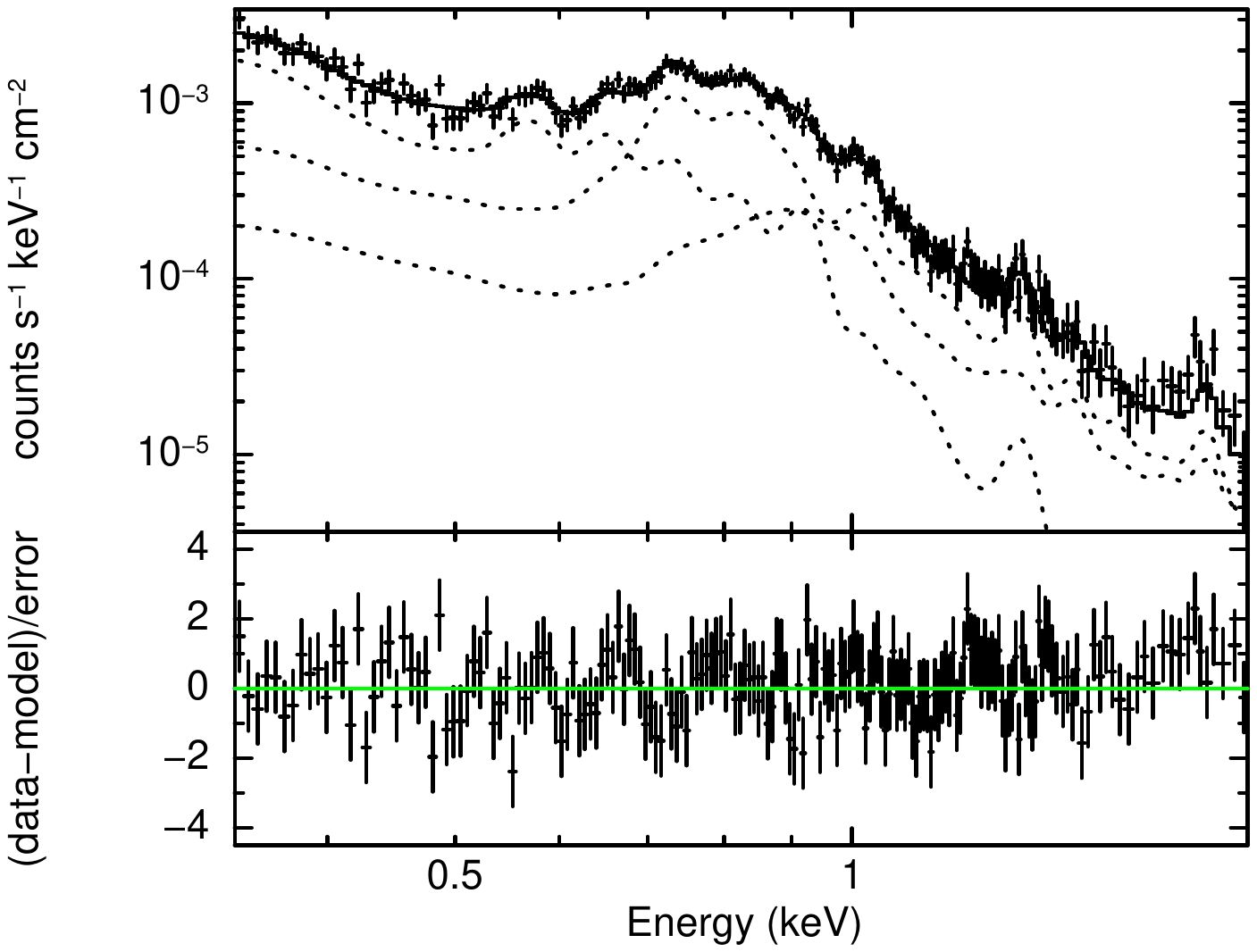}
\hspace{-1cm}
\includegraphics[width=0.495\textwidth]{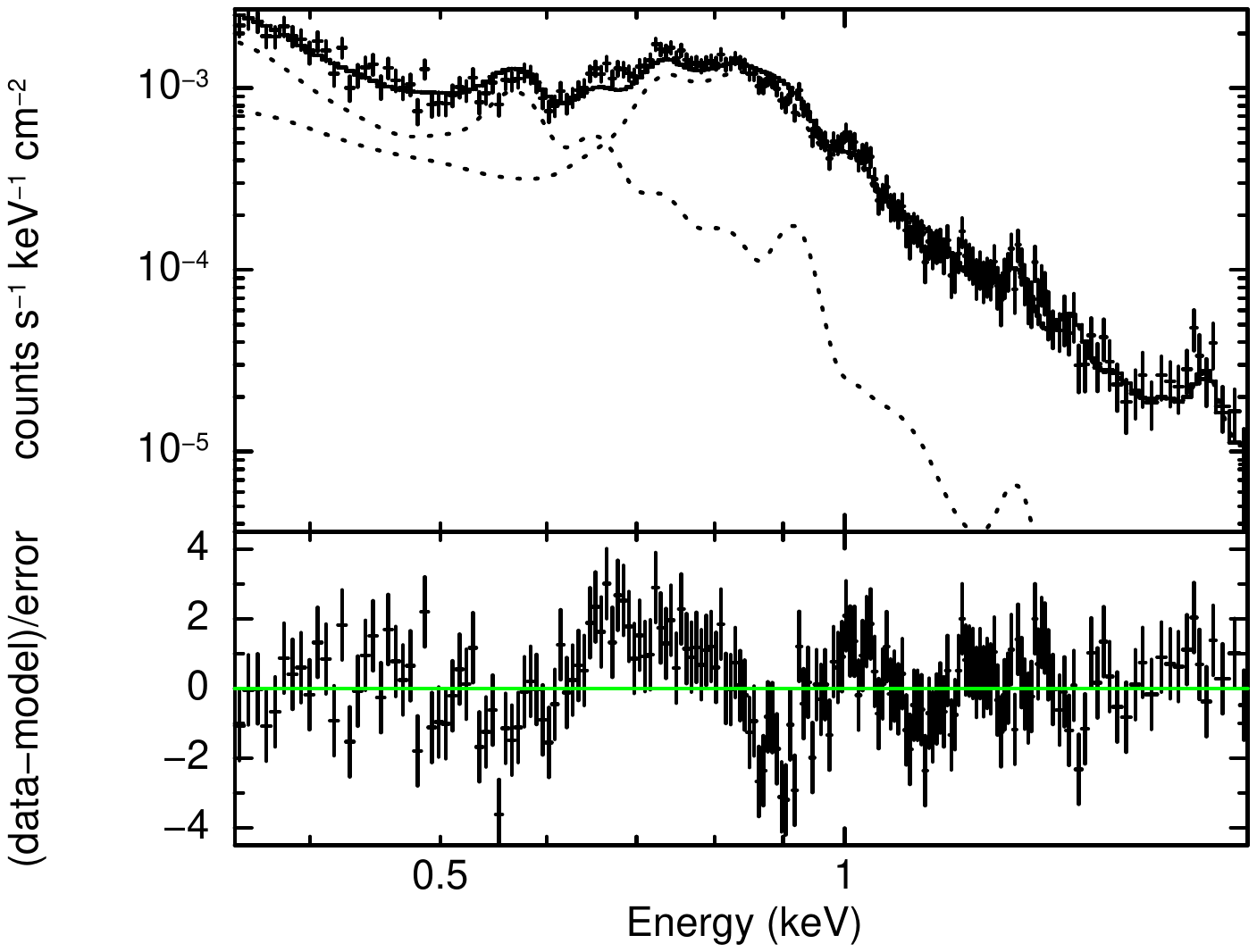}
\vspace{-0.2cm}
\caption{Stacked \erosita\ spectra of the 96 M dwarf (top row) and the 30 FGK stars (bottom row) in the western Galactic hemisphere and within 10~pc of the Sun (Zheng et al. in prep.). The stack is computed as if all stars were at 10~pc from us. 
The left column shows the average spectra and the best fit model, composed of two to three thermal components with variable elemental abundances. In contrast, the right column shows the spectra and residuals for a fit with a two-temperature model with abundances fixed at the solar value and the temperatures fixed at 0.2 and 0.7~keV. }
\label{fig:stars}
\end{figure*}

The top left panel of Fig. \ref{fig:stars} shows the stacked \erosita\ spectrum of the 103 M dwarf stars within the 10~pc sample located within the western Galactic hemisphere (Zheng et al. in prep). 
The distance of each star is known, therefore the stack has been computed after renormalising the flux as if all stars were located at 10~pc from the Sun. 
The spectrum has been fitted with two components of collisionally ionised and optically thin emitting plasma, with variable abundances ({\sc vapec} model in {\sc xspec}). 

The two components have best fit temperatures of $kT_1\sim0.26$ and $kT_2\sim0.95$, and abundances close to 0.1~Solar, apart from Neon, Silicon and Iron that have values of $\sim0.2$~Solar and Oxygen of $\sim0.2$~Solar (see Zheng et al. in prep. for more details). 
Figure \ref{fig:stars} confirms that the spectrum of M dwarfs shows a non-negligible contribution in the 0.7-1.0~keV band. 
To evaluate what could be the contribution of M dwarf stars to the 0.7 keV emission, we fit the stacked \erosita\ spectrum with two components with temperatures fixed at $kT_1=0.2$ and $kT_2=0.7$~keV, as we assumed in our model of the X-ray background. 
Although this model is a poorer fit to the data (top right panel of Fig. \ref{fig:stars}) than the one with free temperatures (top left panel of Fig. \ref{fig:stars}), it allows us more easily to assess the luminosity of M dwarfs at 0.7~keV by simply measuring the flux of the hottest of the two spectral components. 
We find that the 0.7~keV component has a flux of $F_{0.1-10}= 1.6\times10^{-13}$ erg cm$^{-2}$ s$^{-1}$ in the 0.1-10.0 keV band or $L_{0.1-10} = 1.9\times10^{27}$ erg s$^{-1}$. 

To estimate the representativeness of this average luminosity for the M dwarf sample, we take the \cite{Caramazza23} sample, which has determined the luminosity of all the M0-M4 stars within the 10~pc sample. 
We observe that the average luminosity within the eastern Galactic hemisphere is $\sim4.7$ times larger than in the western hemisphere we studied. 
In fact, by chance, the five most luminous stars of the 10~pc M dwarf sample all fall in the eastern Galactic hemisphere, resulting in a higher average luminosity (by a factor of $\sim4.7$) in the eastern, compared with the western hemisphere. 
Therefore, this suggests that the average luminosity of the M dwarf sample might be larger than the one estimated above, with luminosities of the order of $L_{0.1-10} = 5.4\times10^{27}$ erg s$^{-1}$. 

We repeated the same analysis for the FGK stars. 
The bottom left panel of Fig. \ref{fig:stars} shows the stacked \erosita\ spectrum of the 21 FGK stars in the 10~pc sample located within the western Galactic hemisphere (Zheng et al. in prep). 
The spectrum is best fit by three temperature components with temperatures of $kT_1\sim0.23$, $kT_2\sim0.56$ and $kT_3\sim0.9$~keV, respectively, and abundances of about $\sim0.3$~Solar, (apart from Oxygen, Neon, Silicon and Iron that have values of $\sim0.2$, $0.6$, $0.1$ and $0.35$~Solar, respectively). 
To estimate the contribution of the FGK stars to the 0.7~keV emission observed at 0.7~keV, we re-fit the spectrum with a two-temperature component, fixing the temperatures to $kT_1=0.2$ and $kT_2=0.7$~keV (see bottom right panel of Fig. \ref{fig:stars}). 
The 0.7~keV component has a flux of $F_{0.1-10}= 9.0\times10^{-13}$ erg cm$^{-2}$ s$^{-1}$ in the 0.1-10.0 keV band or $L_{0.1-10} = 1.07\times10^{28}$ erg s$^{-1}$. 

We note the importance of evaluating the stellar contribution to the Galactic X-ray emission on a volume-complete sample, which probes the full range of X-ray luminosities that occur in nature. 
If we had neglected X-ray undetected stars, we would have overestimated the average stellar X-ray luminosity per
mass. 
On the other hand, flux-limited samples are biased towards the stars with the highest X-ray luminosities, which provide the strongest stellar contribution to the large-scale 0.7~keV emission. 
Stars emitting at the X-ray saturation level, $L_{\rm x,sat}$, which is a function of stellar mass, are rare in the $10$\,pc sample (\cite{Caramazza23}, Bennedik et al., in prep.). 
However, such highly active stars are rather abundant within a few hundred parsecs of the Sun. 
Indeed, we find that young stars in nearby star-forming regions significantly contribute to the observed 0.7~keV emission. 

\section{Removed bright point sources and its stars fraction}
\label{sec:starFrac}

\begin{figure}[t]
\centering
\includegraphics[width=0.39\textwidth]{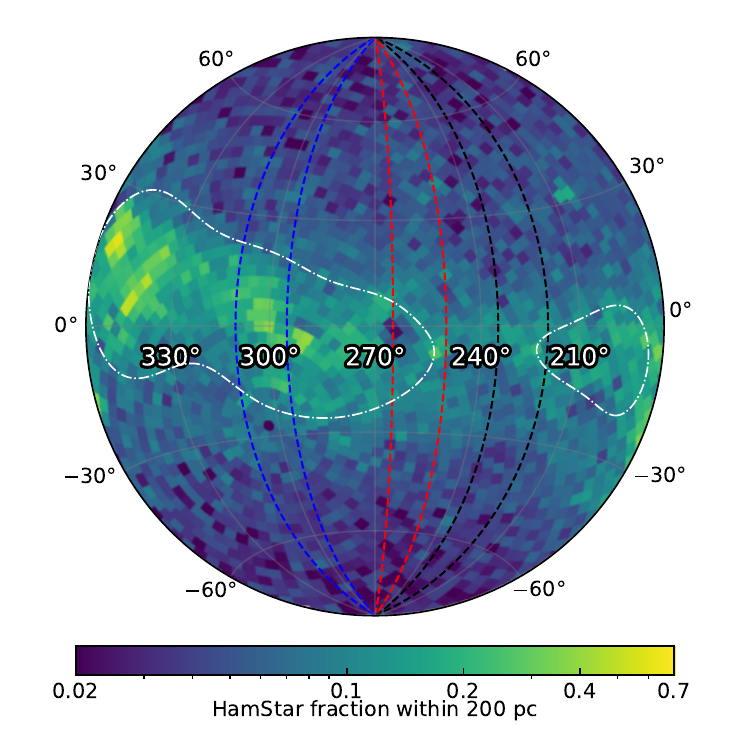}
\vspace{-0.3cm}

\includegraphics[width=0.39\textwidth]{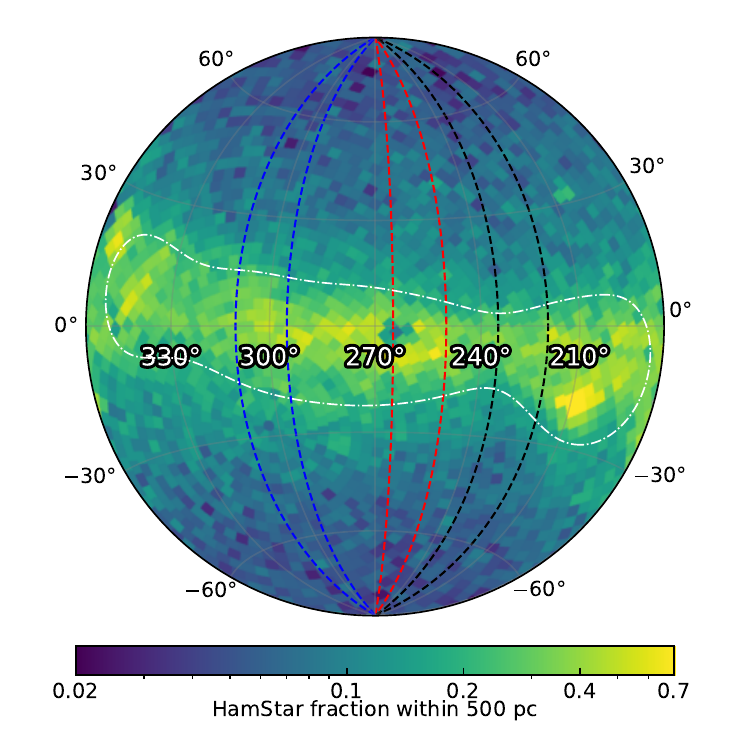}
\vspace{-0.36
cm}

\includegraphics[width=0.39\textwidth]{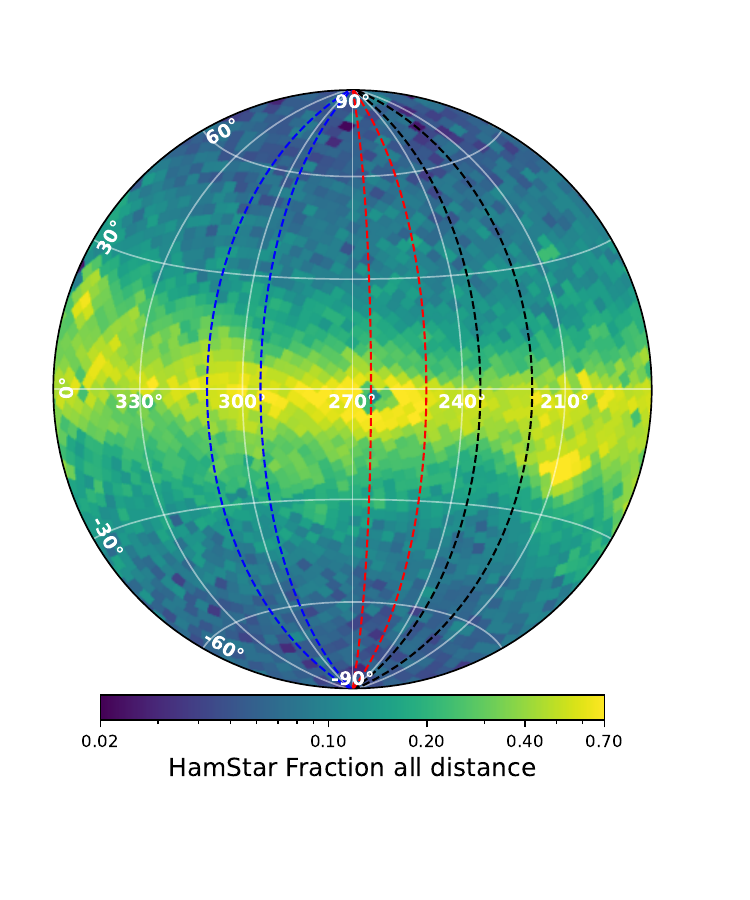}
\vspace{-1.3cm}

\caption{Fraction of bright point sources identified as stars by \cite{Freund24}. 
{\it Top and middle panels:} Fraction of stars within 200~pc and 500~pc, respectively. 
The fraction of stars within the solar neighbourhood is asymmetric with respect to the Galactic plane, with fractions as high as 40--70~\% along the nearby stellar structures. The dot-dashed white contours indicate the region with the highest concentration of stars within 200~pc (top) and 500~pc (middle) of the Sun. They are derived from the respective smoothed images. Such as the 0.7~keV emission, the profile of the fraction of detected stars within 500~pc of the Sun is skewed towards the south at $220^\circ<l<235^\circ$ and $250^\circ<l<265^\circ$, while it is nearly symmetric at $295^\circ<l<310^\circ$. 
{\it Bottom panel:} Fraction of stars at any distance from the Sun.} 
\label{fig:StarFrac}
\end{figure}

Figure \ref{fig:StarFrac} shows which fraction of the bright sources is classified as a star by the HamStar catalogue \citep{Freund24}. 
Figure \ref{fig:StarFrac} clearly shows that the fraction of X-ray bright stars within few hundred parsecs of the Sun is not perfectly symmetric around the Galactic plane and the characteristic shape of the nearby young stellar structures appears evident \citep{Gould79,Guillout98,Alves20,Zucker22,Freund24}. 

The bottom panel of Fig. \ref{fig:StarFrac} shows all stars, including the ones beyond 500~pc. 
As expected, including the more distant stars greatly increases the star fraction along the Galactic plane.
Indeed, stars compose up to 70\% of all bright point sources close to the Galactic plane, while at high Galactic latitudes the fraction drops to about 15\%. 

\section{Distance versus mass distribution}

\begin{figure}[h]
\centering
\includegraphics[width=0.49\textwidth]{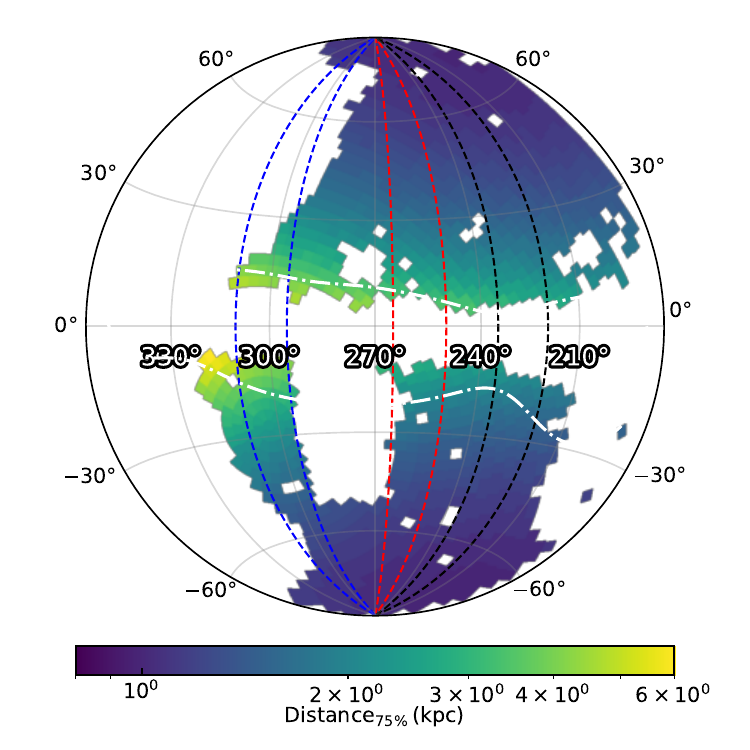}
\caption{Distance along the line of sight within which 75~\% of the stellar mass is contained, assuming the \cite{Hunter24} stellar mass distribution. 
} 
\label{fig:Dist}
\end{figure}

Figure \ref{fig:Dist} shows the distance from the Sun at which the integral of the stellar mass distribution (eq.~\ref{eq:sigma_M}) reaches 75~\% of its total projected value along a given line of sight, computed for all lines of sight in the Western sky. 
More than 75~\% of the mass is contained within 1~kpc, at high Galactic latitudes, tracing stellar populations relatively close to the Sun. 
Closer to the Galactic disc, the stellar mass extends beyond 1 kpc, with values as large as 2--3 kpc and reaching 75~\% of the mass within 4--6~kpc close to the disc and closer to the Galactic centre. 

Considering the patchy distribution of the cold interstellar medium, it is likely that several absorption layers are affecting the spectra along the lines of sight where the emission extends to larger distances. 
This makes the analysis ambiguous because it cannot be established a priori how the emission and absorption are distributed along the line of sight. 
To avoid these complications, we have selected only sky tiles for which $\log(N_{{\rm H, HI4}})<21.5$, a value for which the 0.7~keV component is close to transparent. 

\end{appendix}

\end{document}